\documentclass{ws-jca}

\usepackage{graphicx}

\begin{document}

\newcommand{\F}{_{\mathrm{f}}}
\newcommand{\B}{_{\mathrm{b}}}

\markboth{D.~Makarov}
{Random matrix theory for low-frequency sound propagation in the ocean: a spectral statistics test}

%
\catchline{}{}{}{}{}
%

\title{RANDOM MATRIX THEORY FOR LOW-FREQUENCY SOUND PROPAGATION IN THE OCEAN: \\
A SPECTRAL STATISTICS TEST}

\author{Denis Makarov}

\address{Laboratory of Nonlinear Dynamical Systems,\\ V.I.Il'ichev Pacific Oceanological Institute of FEB RAS ,\\
690041, Baltiyskaya Str. 43, Vladivostok, Russia \\
\email{makarov@poi.dvo.ru} 
\http{$<dynalab.poi.dvo.ru>$} 
}



\maketitle

\begin{history}
\received{(Day Month Year)}
\revised{(Day Month Year)}
\end{history}

\begin{abstract}
Problem of long-range sound propagation in the randomly-inhomogeneous deep ocean is considered.
We examine a novel approach for modeling of wave propagation, developed by K.C. Hegewisch and S.Tomsovic.
This approach relies on construction of a wavefield propagator using the random matrix theory (RMT).
We study the ability of the RMT-based propagator to reproduce properties of the propagator
corresponding to direct numerical solution of the parabolic equation.
It is shown that mode coupling described by the RMT-based propagator is basically consistent with the direct Monte-Carlo simulation.
The agreement is worsened only for relatively short distances, when long-lasting cross-mode correlations 
are significant. It is shown that the RMT-based propagator with properly chosen range step can reproduce
some coherent features in spectral statistics.

\end{abstract}


\section{Introduction}
\label{Intro}

Development of efficient and computationally cheap methods 
for modeling of long-range sound propagation in randomly inhomogeneous waveguides
is one of the most important problems of ocean acoustics.
Indeed, even weak spatial variability of the sound speed,
caused by oceanic internal waves, becomes a substantial factor
on the ranges of few hundred kilometers \cite{AET,RayWave,UFN,ColosiBook}. 
Therefore, calculations with range-independent or adiabatically varying
models of medium couldn't reproduce actual properties of sound propagation.
On the other hand,
acoustical experiments are commonly conducted with pulse broadband signals.
In this way, the most simple approach based on the synthesis of an acoustic pulse from monochromatic components, 
together with Monte-Carlo sampling, 
can be excessively cumbersome.
To avoid the above difficulties, one can utilize the kinetic approach based on construction of the master equation
for mode amplitudes of a wavefield \cite{ColosiBook,Dozier_Tappert,ColosiMorozov}. 
This approach can be used for studying spatial and temporal coherences of a wavefield \cite{CCVO,Viro-WRCM16}.

Another promising approach is 
modeling of sound propagation by means of the random matrix theory (or shortly RMT) \cite{Hege-EPL,Hege-JASA13}.
Its important advantage is the ability to model interference patterns and acoustic timefronts without solving the wave equation directly.
In the RMT approach, complex-valued amplitudes of inter-mode transitions are considered 
as statistically independent random quantities, with
variances determined by the random inhomogeneity of a waveguide.
In this way, one basically ignores correlations between different transitions. 
These correlations, however, might be responsible 
for various coherent phenomena that can be observed in experiments. 
For example, they can give rise to regularly propagating beams formed by coherent ray clusters \cite{Chaos}.
As it was shown in Ref.~\cite{PRE87},
coherent features are reflected in 
 spectral properties of the propagator governing wave evolution.
So, one can ask: whether the RMT approach is able to reproduce actual properties of sound propagation?
This is namely the issue we address in the present work. We compare results obtained with the random matrix theory
with those obtained by solving the parabolic equation. 
We restrict our attention by long-range sound propagation in the deep ocean. In particular, we use the
canonical Munk waveguide and the waveguide with the biexponential sound-speed profile \cite{RayWave,Chaos}. Both models include random inhomogeneity
caused by internal waves.

The manuscript is organized as follows.
A wavefield propagator is introduced in the next section.
Section \ref{RMT} describes construction of the propagator by means of the random matrix theory.
Section \ref{Models} is devoted to models of an underwater sound channel used in numerical simulation.
In Section \ref{Numer0} we examine the ability of the RMT approach to describe dynamics of modal amplitudes.
Some fundamental features of spectral analysis of the propagator are described in Section \ref{Spectrum}.
Results of numerical simulation of spectral statistics are presented in Sections \ref{Numer}.
In Summary, we give the main conclusions and outline perspectives of the future research.

\section{Unitary propagator for sound propagation in the ocean and its spectral properties}
\label{Operator}


Consider an underwater sound channel where spatial variability
of sound speed is represented by the sum
\begin{equation}
 c(z,\,r)=c_0+\Delta c(z)+\delta c(z,\,r),
\label{c}
\end{equation}
where $z$ is depth, $r$ is the range coordinate,
$c_0$ is a reference sound speed, 
$\Delta c(z)$ corresponds to a background range-independent sound-speed profile,
and $\delta c(r,z)$ is random sound-speed inhomogeneity caused by oceanic internal waves.
Terms entering into Eq.~(\ref{c}) satisfy the inequality
\begin{equation}
\lvert\delta c\rvert_\text{max}\ll \lvert\Delta c\rvert_\text{max}\ll c_0.
\label{ocean}
\end{equation}
Owing to this inequality, we can invoke the small-angle approximation, 
when an acoustic wavefield is governed by the standard parabolic
equation
\begin{equation}
\frac{i}{k_0}\frac{\partial\Psi}{\partial r}=
-\frac{1}{2k_0^2}\frac{\partial^2\Psi}{\partial
z^2}+\left[U(z)+V(r,\,z)\right]\Psi, 
\label{parabolic}
\end{equation}
where 
wave function $\Psi$ is related to
acoustic pressure $u$ by means of the formula $u=\Psi\exp(ik_0r)/\sqrt{r}$. 
Quantity $k_0$ is a reference wavenumber related to sound
frequency $f$ as 
\begin{equation}
 k_0=\frac{2\pi f}{c_0}.
 \label{k0}
\end{equation}
Functions $U(z)$ and $V(r,z)$ are determined by spatial sound-speed variations.
In the the small-angle approximation they can be expressed as
\begin{equation}
U(z)=\frac{\Delta c(z)}{c_0},\quad 
V(r,\,z)=\frac{\delta c(r,\,z)}{c_0}.
\label{pot}
\end{equation}
In the present paper we use idealistic perfectly-reflecting boundary conditions of the form
\begin{equation}
\left.\Psi\right\vert_{z=0} = 0,\quad
\left.\frac{d\Psi}{dz}\right\vert_{z=h}=0.
\label{BCs}
\end{equation}
Solution of the parabolic equation (\ref{parabolic}) can be represented as a sum over normal modes
\begin{equation}
 \Psi(r,z,k_0) = \sum\limits_{m} a_m(r,k_0)\psi_m(z)
\end{equation}
The normal modes and the corresponding eigenvalues satisfy the Sturm-Liouville problem
\begin{equation}
-\frac{1}{2k_0^2}\frac{\partial^2\psi_m(z)}{\partial
 z^2}+U(z)\psi_m(z)=E_m\psi_m(z).
\label{StL}
\end{equation}
In a range-independent waveguide ($\delta c = 0$) we have
\begin{equation}
 a_m(r,k_0) = a_m(r=0,k_0)e^{-ik_0E_mr},
\end{equation}
i.~e. modes are uncoupled and
there is no transfer between different modes.
Horizontal inhomogeneity of a waveguide leads to mode coupling.

As long as the boundary conditions (\ref{BCs}) provide preservation of the wavefunction norm,
general solution of the parabolic equation at range $r=r\F$
can be formally written in terms of an unitary propagator $\hat G$ \cite{Froufe-Perez} acting as
\begin{equation}
 \Psi(r\F,z) = \hat G(r_0,r\F)\Psi(r_0,z).
 \label{evolution}
\end{equation}
In fact, unitarity (i.~e. norm preservation) of the propagator is provided by somewhat artificial 
assumptions.
Indeed, the norm preservation means that we neglect sound absorption within the ocean bottom and the water volume.
In the case of the deep ocean, interaction with the bottom doesn't affect a large group of the lowest modes
which are responsible for long-range propagation. As about the water volume attenuation, it is fairly weak for frequencies below 100 Hz
and doesn't influence process of sound scattering.
Therefore, the approximation of the unitary evolution is quite reasonable for our purposes.

The propagator $\hat G$ can be represented in the matrix form using the basis of normal modes (\ref{StL}).
The corresponding matrix elements are 
\begin{equation}
 G_{mn}(r_0,r\F)=\int \psi_m^*\hat G(r_0,r\F)\psi_n\,dz,
 \label{Gmn}
\end{equation}
where $\hat G(r_0,r\F)\psi_n$ is a solution of the parabolic equation at range $r=r\F$
for the initial condition $\Psi(r=r_0)=\psi_n$.
Since $G_{mn}=G_{nm}^*$ for $m\ne n$, the matrix of the propagator is Hermitian.

\section{Construction of the propagator by means of the random matrix theory}
\label{RMT}

In the present section we give brief description of the random matrix approach proposed in Refs.~\cite{Hege-EPL,Hege-JASA13}.
As a starting point, we express the propagator $G(r_0,r\F)$ as a product of propagators for intermediate segments of a waveguide:
\begin{equation}
 \hat G(r_0,r_J) = \hat G(r_{J-1},r_J)\hat G(r_{J-2},r_{J-1}\text)\text{...}\hat G(r_{1},r_{2})\hat G(r_{0},r_{1}),
 \quad
 r_J=r\F.
 \label{product}
\end{equation}
Hereafter we set $r_0=0$ and consider the equally spaced partition of the interval $[0:r\F]$.
The spacing $\Delta r=r_{j}-r_{j-1}$ can be chosen large enough to provide statistical independence of propagators for neighboring segments.
It is reasonable to assume that $\delta c(r)$ is a stationary stochastic process.
Then we can replace $\hat G(r_{j-1},r_j)$ by $\hat G(\Delta r)$ and rewrite Eq.~(\ref{product}) as 
\begin{equation}
\hat G(0,r\F)=\hat G(J\Delta r)=\prod\limits_{j=1}^{J}\hat G_j(\Delta r),
 \label{BB}
\end{equation}
where the index $j$ enumerates propagators corresponding to different realizations of $\delta c$.

To construct a realization of a single-segment propagator, it is reasonable to utilize the matrix representation (\ref{Gmn}).
Then we can replace the propagators $\hat G$ in Eqs. (\ref{product}) and (\ref{BB}) by the corresponding matrices $\mathbf{G}$.
The first-order perturbation theory yields
\begin{equation}
 \mathbf{G}(\Delta r) = \mathbf{\Lambda}(\mathbf{I} - i\mathbf{A}),
 \label{1order}
 \end{equation}
where $\mathbf{I}$ is the identity matrix, and $\mathbf{\Lambda}$ is a diagonal matrix with elements
\begin{equation} 
 \Lambda_{mn} = \delta_{mn}e^{-ik_0E_m\Delta r},
 \end{equation} 
where $\delta_{mn}$ is the Kronecker symbol. 
In Eq.~\ref{1order}, $\mathbf{A}$ is an inhomogeneity-induced perturbation matrix whose elements are calculated as
 \begin{equation}
  A_{mn}=k_0e^{ik_0(E_m-E_n)r'}\int\limits_{r'=0}^{\Delta r}  V_{mn}(r')\,dr',
 \label{pert}
\end{equation}
\begin{equation}
V_{mn}(r) = \int \psi_m^*(z)V(r,z)\psi_n(z)\,dz.
 \label{Vmn}
\end{equation}
The key idea of the random matrix approach is to treat matrix elements of the perturbation $\mathbf{A}$ as 
random quantities
\begin{equation}
 A_{mn}(\Delta r,k_0) = \sigma_{mn}(\Delta r,k_0)z_{mn}(k_0),
 \label{Amn}
\end{equation}
where $\sigma_{mn}$ is calculated from spectral properties of the random inhomogeneity,
and $z_{mn}$ is a complex-valued Gaussian random variable with the unit variance.

The first-order perturbation theory basically doesn't maintain unitarity.
To restore unitarity without altering significantly physical properties, authors of Refs. \cite{Hege-EPL,Hege-JASA13} 
offer to use the Cayley transform 
\begin{equation}
\mathbf{G}(\Delta r) = \mathbf{\Lambda}[\mathbf{I} + i\mathbf{A}(\Delta r)/2]^{-1}[\mathbf{I} - i\mathbf{A}(\Delta r)/2]
\label{Cayley} 
\end{equation}
In this case the matrix $\mathbf{G}$ is guaranteed to be unitary.

Mode amplitudes of a wavefield can be combined into the vector $\vec{a}$, $\vec{a}\equiv (a_1, a_2, \text{...}, a_M)^T$.
In accordance with Eq.~(\ref{evolution}), range evolution of this vector is governed by the equation
\begin{equation}
 \vec{a}(r) = \mathbf{G}(r)\vec{a}(0).
 \label{amod}
\end{equation}
So, a wavefield can be calculated by means of sequential multiplication of the vector of mode amplitudes by the propagator matrix.
This algorithm is extremely fast if the matrix size (i.~e. number of trapped modes) is not very large.
Indeed, complexity of multiplication of a matrix $\mathbf{G}$ by the vector $\vec{a}$ scales as $M^2$, where $M$ is number of trapped modes.

\section{Models of an underwater sound channel}
\label{Models}

\begin{figure}[!htb]
\begin{center}
\includegraphics[width=0.6\textwidth]{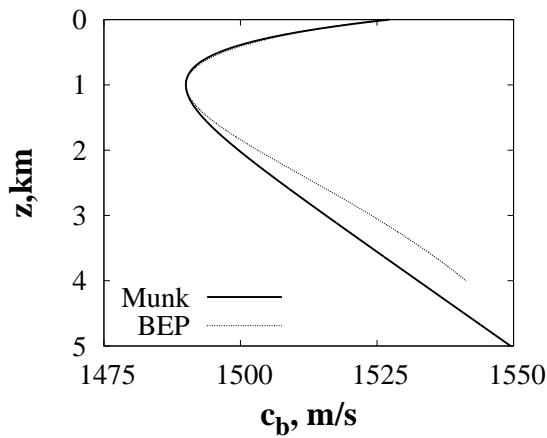}
\caption{Background sound-speed profiles corresponding to the canonical Munk 
and biexponential (BEP) models.
}%
\label{fig-prof}
\end{center}
\end{figure}

In this work we consider two models of an underwater sound channel.
The first one is the so-called biexponential (or shortly BEP) model  originally introduced in Ref.~\cite{DAN}
and further developed in Refs.~\cite{RayWave,Chaos}. In the BEP model, the background 
sound-speed profile is determined by equation
\begin{equation}
 c\B(z) = c_0 + \Delta c(z) = c_0\left[
 1+\frac{b^2}{2}(e^{-az} - \kappa)^2 \right],
 \label{BEP}
\end{equation}
Here $c_0=1490$~m/s, $a=0.5$~km$^{-1}$, $b=0.557$, and $\kappa=0.6065$. Ocean bottom is located at the depth of 4 km.
An important advantage of the BEP model is the presence of analytical expressions for some important waveguide
characteristics, like ray cycle length, ray travel times, e.~t.~c. \cite{RayWave,Chaos}

The second considered model is the celebrated canonical Munk waveguide with
the background sound-speed profile
\begin{equation}
 c_{\text{b}}(z) =c_0\left[
 1 + \gamma (e^{-\eta}-1+\eta)
 \right],\quad \eta = \frac{2(z-z_{\text{a}})}{B},
\end{equation}
where $c_0=1490$~m/s, $B=1$~km, $\gamma=0.0057$, and the ocean bottom is located at $z=5$~km. Both sound-speed profiles are shown in Fig.~\ref{fig-prof}.

\begin{figure}[!htb]
\begin{center}
\includegraphics[width=0.9\textwidth]{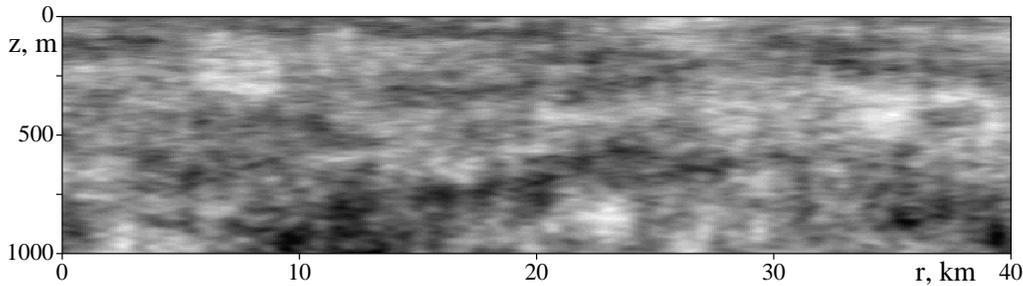}
\caption{Field of vertical displacements of a fluid parcel, the upper part, a single realization. 
The brightest light and dark colours correspond to displacements of -15 and 15 m, respectively.
}%
\label{fig-zeta}
\end{center}
\end{figure}

Sound-speed perturbation induced by internal waves is constructed according to the scheme presented in Refs.~\cite{Hege-JASA13,ColBr}.
In that scheme,
one assumes linear dependence of sound-speed variations $\delta c$ on vertical displacement of a fluid parcel $\zeta$
\begin{equation}
 \delta c(r,z) = c_0\left(\frac{24.5}{g}\right)N^2\zeta,
\end{equation}
where $g=9.8$ ms$^{-2}$ is the gravitational acceleration, and $N$ is V\"ais\"ala-Brunt frequency.
Wavefield of $\zeta$ is given by the double sum
\begin{equation}
 \zeta(r,z) = \frac{2B}{\pi}\sqrt{\frac{E\Delta k_l}{M}}e^{-3z/2B}\sum\limits_{j=1}^{j_{\max}}\sum\limits_{l=1}^{l_{\max}}
 \sin(j\pi\xi(z))\sqrt{\frac{I_{j,k_l}}{j^2+j_*^2}}\cos(k_lr+\phi_{jl}),
 \end{equation}
where $\xi(z)=e^{-z/B}-e^{-h/B}$, and $\phi_{jl}$ are random phases uniformly distributed over the interval $[0:2\pi]$. Spectral weights are described by formula
\begin{equation}
I(j,k_l) = \frac{k_j}{k_l^2+k_j^2} + \frac{1}{2}\frac{k_l^2}{(k_l^2+k_j^2)^{3/2}}
\text{ln}\frac{\sqrt{k_l^2+k_j^2}+k_j}{\sqrt{k_l^2+k_j^2}-k_j},
 \label{GM}
\end{equation}
where
\begin{equation}
 k_j = \frac{\pi jf_{\text{i}}}{N_0B}.
\end{equation}
The following values of parameters are taken:
$N_0=2\pi/10$ min, $f_{\text{i}}=1$ cycle per day, the Garrett-Munk energy $E=6.3*10^{-5}$, mode scaling number $M=(\pi j_*-1)/2j_*^2$, 
and the principle mode number $j_*=3$.
We take 1000 values of
the horizontal internal wave number $k_l$, which are equally spaced within the interval from $k_l=2\pi/100$ to $2\pi$ 
radians per km, with spacing $\Delta k_l$. 
A single realization of $\zeta(r,z)$ is plotted in Fig.~\ref{fig-zeta}.

Construction of a proper random matrix ensemble 
for the propagator $\hat G$
can be further facilitated 
by expanding the inhomogeneity over empirical orthogonal functions \cite{Radiophys,LeBlanc}
\begin{equation}
 \delta c_\mathrm{iw}(z,r)=<\delta c_\mathrm{iw}(z)>+\sum\limits_k q_k(r)\theta_k(z).
\label{KL}
\end{equation}
This expansion is also known as the Karhunen-Loeve expansion.
Empirical orthogonal functions $\theta_n(z)$ are the eigenvectors of the covariance matrix
$\mathbf{K}$ whose elements are given by
\begin{equation}
K_{ij}=\frac{1}{l_{\text{max}}}\sum\limits_{l=1}^{l_{\text{max}}}[\delta c_{l}(z_i)
-<{\delta c}(z_i)>]*[\delta c_{l}(z_j)-<{\delta c}(z_j)>],
 \label{covar}
\end{equation}
where the summation goes over $l_{\text{max}}$ statistically independent realizations of 
и $\delta c(z)$, $\{z_i\}$ is a vector of depth values,
and angular brackets denote statistical averaging. 
Taking into account that $\delta c$ is an oscillating function,
we can set $<\delta c_\mathrm{iw}>=0$. 
Eigenvalues of $\mathbf{K}$ quantify contributions of the corresponding eigenvectors in the expansion  (\ref{KL}).
We retain 100 empirical eigenfunctions with the largest eigenvalues that provides almost complete reproduction of an internal wave field.
Then variances in Eq.~(\ref{Amn}) are calculated as
\begin{equation}
\sigma_{mn} = \sum\limits_{k=1}^{N_f} Q_{mn}^{(k)}\sigma_{mn}^{(k)},
\end{equation}
where
\begin{equation}
 Q_{mn}^{(k)} = \frac{1}{c_0}\int \psi_m^*\theta_k(z)\psi_n\,dz,
\end{equation}
and $\sigma_{mn}^{(k)}$ is variance of the integral
\begin{equation}
Y_{mn}^{(k)} = \int\limits_{r=0}^{\Delta r}q_k(r)e^{ik_0(E_m-E_n)r}\,dr.
 \label{eofcoef}
\end{equation}
Utilizing the Karhunen-Loeve expansion, we separate integration over $r$ and $z$ from each other, and thereby
reduce computational efforts to calculate matrix elements $A_{mn}$.

\section{RMT modeling vs direct calculation. I. Mode amplitudes}
\label{Numer0}

\begin{figure}[!htb]
\includegraphics[width=0.48\textwidth]{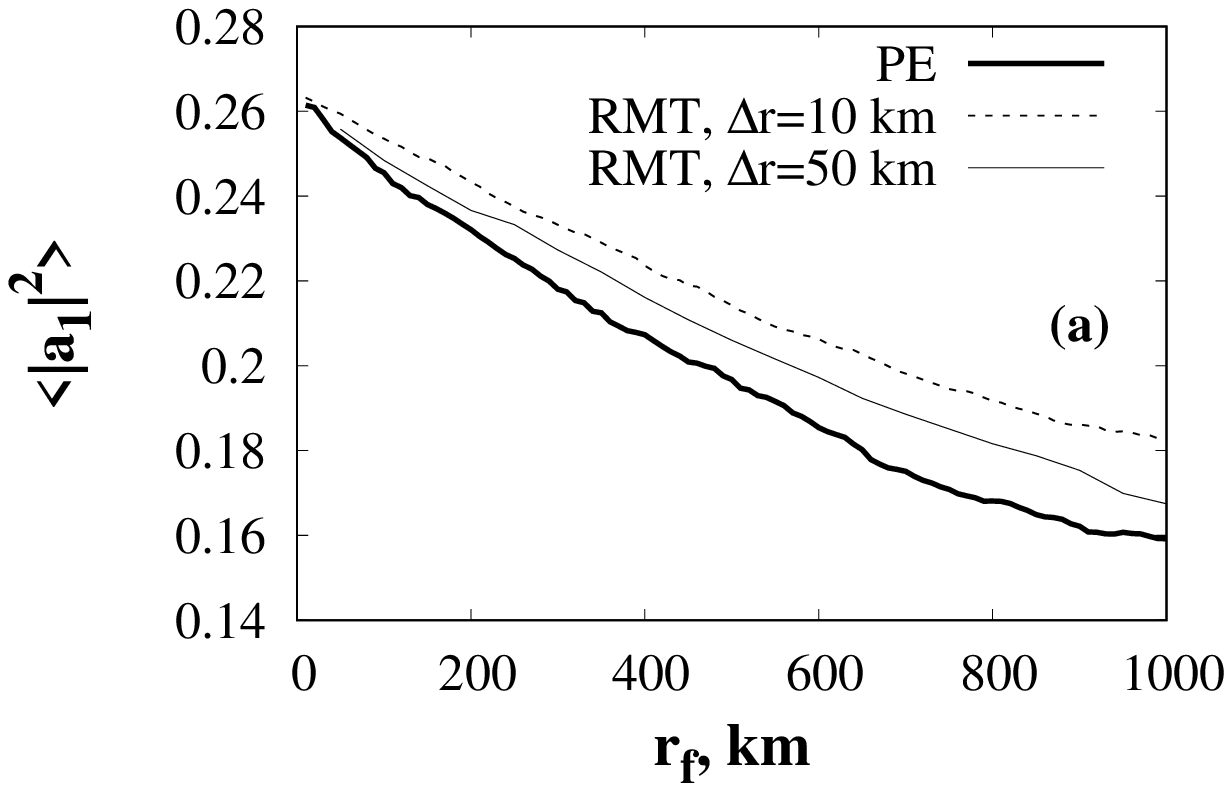}
\includegraphics[width=0.48\textwidth]{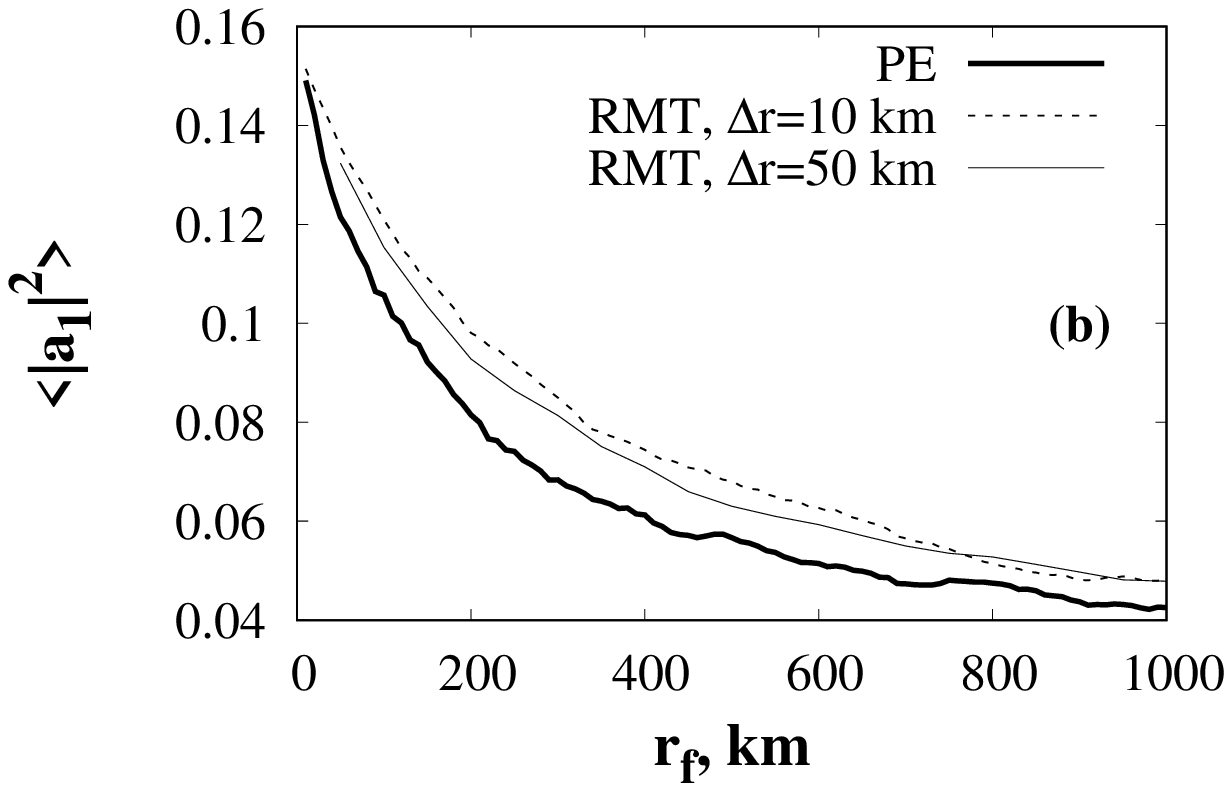}\\
\includegraphics[width=0.48\textwidth]{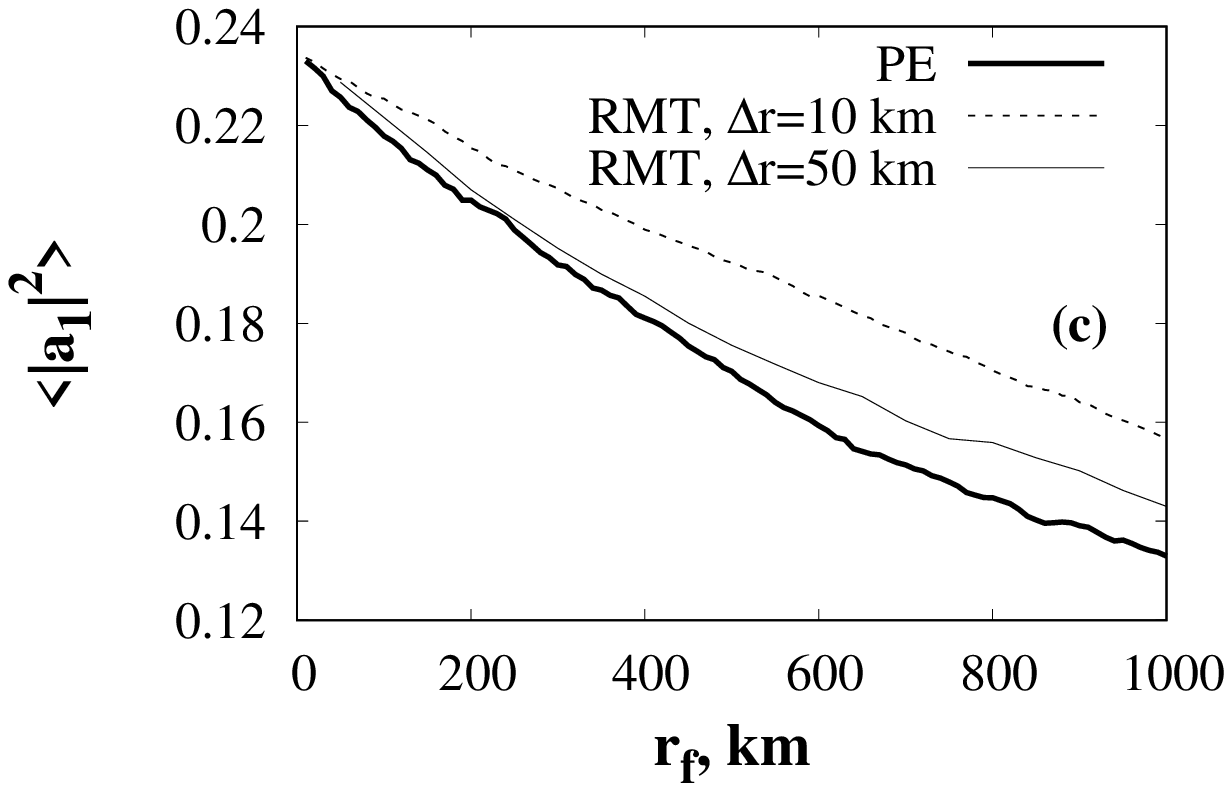}
\includegraphics[width=0.48\textwidth]{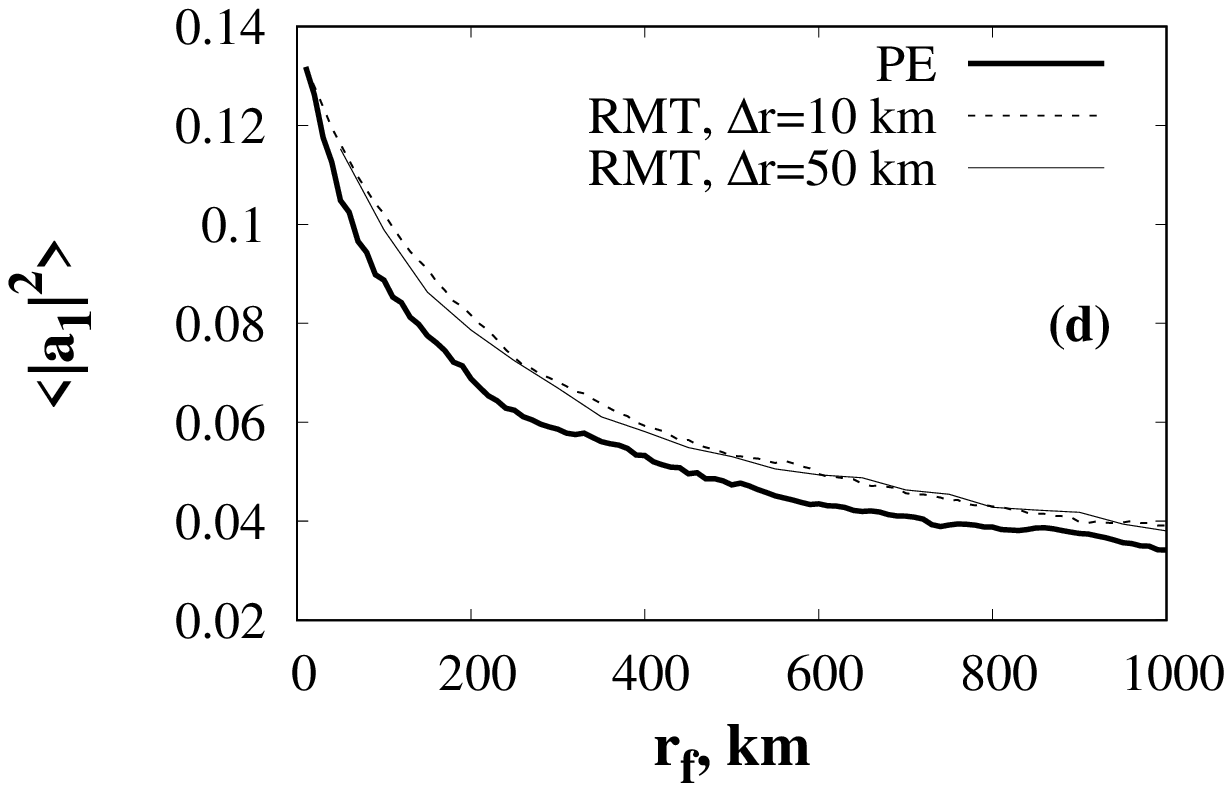}
\caption{Acoustical energy corresponding to the first mode of a waveguide vs range. 
Panels (a) and (b) depict results for the BEP model,
panels (c) and (d) correspond to the Munk model.
Signal frequency: 25 Hz (panels (a) and (c)) and
75 Hz (panels (b) and (d)).
The curves obtained via
the direct solving of the parabolic equation are denoted as ``PE''.
}%
\label{fig-1mode}
\end{figure}

Now let's consider how efficiently the RMT-based propagator reproduces evolution
of the modal spectrum of a wavefield.
We compare results obtained via the RMT with those obtained by means of direct Monte-Carlo sampling.
The direct calculation was performed using the Crank-Nicolson scheme for numerical solution of the parabolic equation.
The RMT modeling was carried out with two values of the propagator step $\Delta r$, 10 and 50 km.
The case of $\Delta r=10$~km corresponds to smaller magnitudes of matrix elements $A_{mn}$, thus providing
better accuracy of the first-order perturbation theory the RMT approach relies upon.
However, it can inaccurately approximate of the actual propagator in the presence of cross-mode correlations whose 
length exceeds 10 km.
From this viewpoint, the case of $\Delta r=50$~km looks as a more reasonable choice.

We consider two values of signal frequency $f$, 25 and 75 Hz. The case of $f=25$~Hz anticipates nearly-adiabatic propagation of modes, with relatively weak inter-mode coupling, 
while in the case of $f=75$~Hz ray-like effects are expected to be relevant.
Ensemble of 1000 realizations was used for statistical averaging. 

\begin{figure}[!htb]
\includegraphics[width=0.48\textwidth]{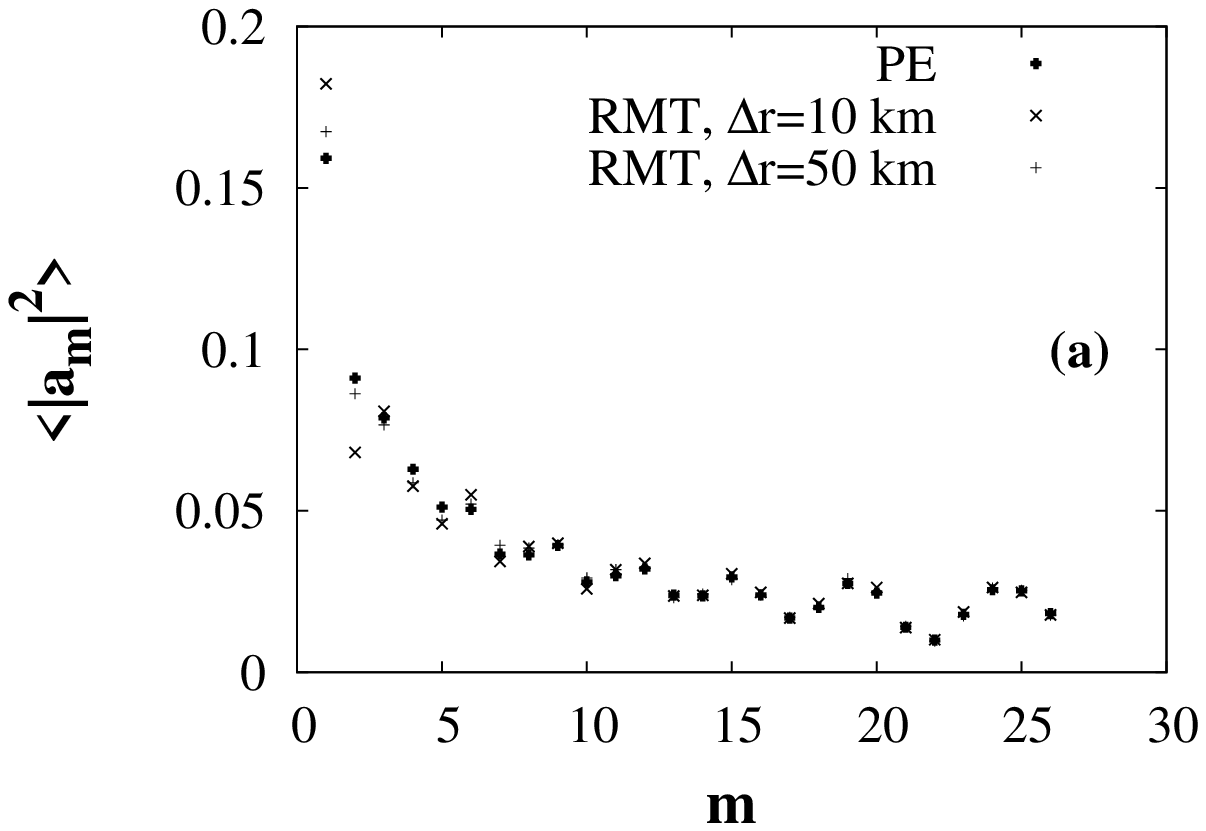}
\includegraphics[width=0.48\textwidth]{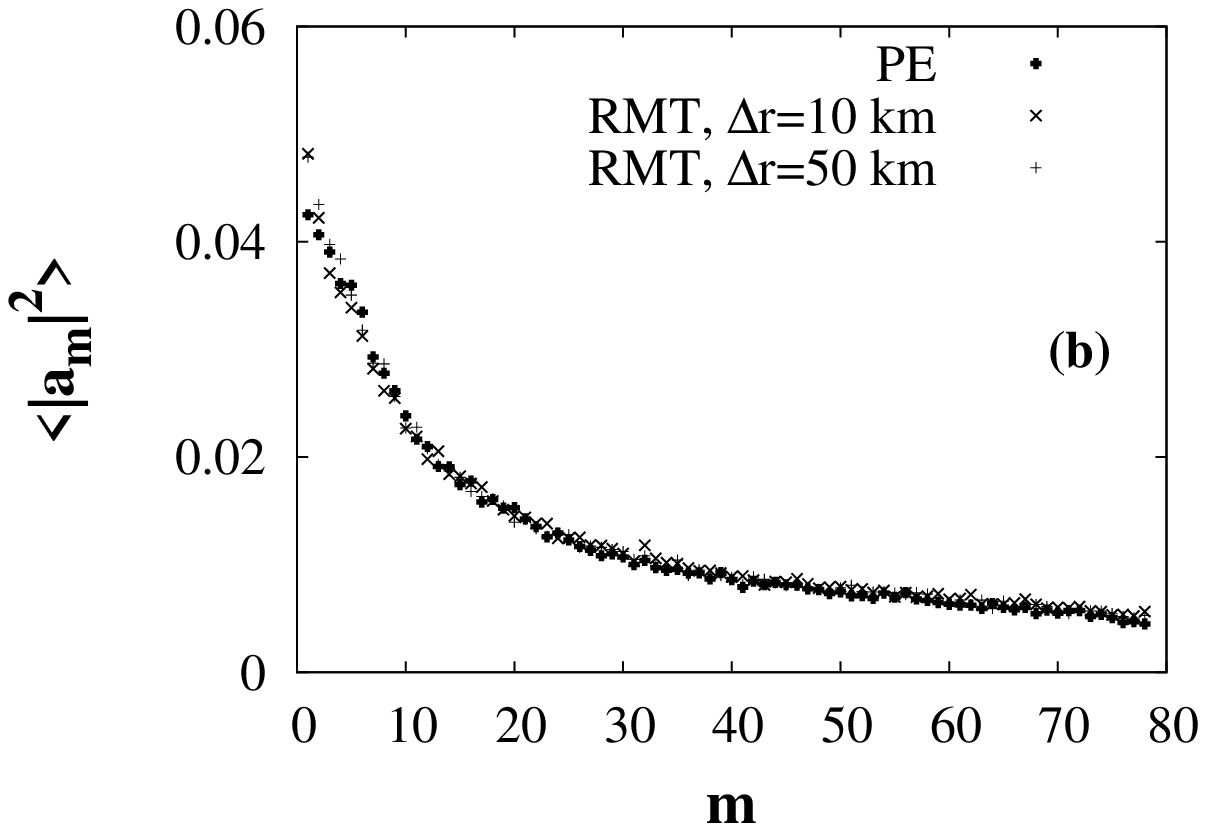}\\
\includegraphics[width=0.48\textwidth]{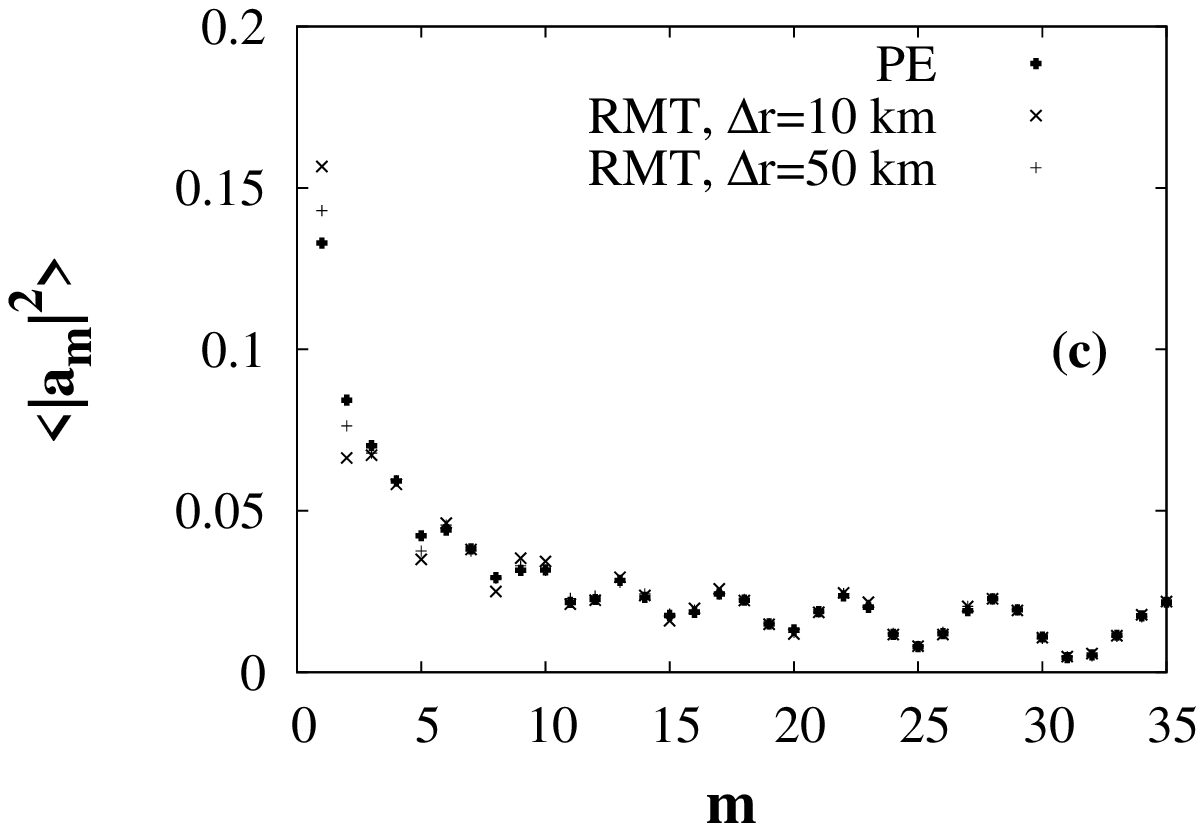}
\includegraphics[width=0.48\textwidth]{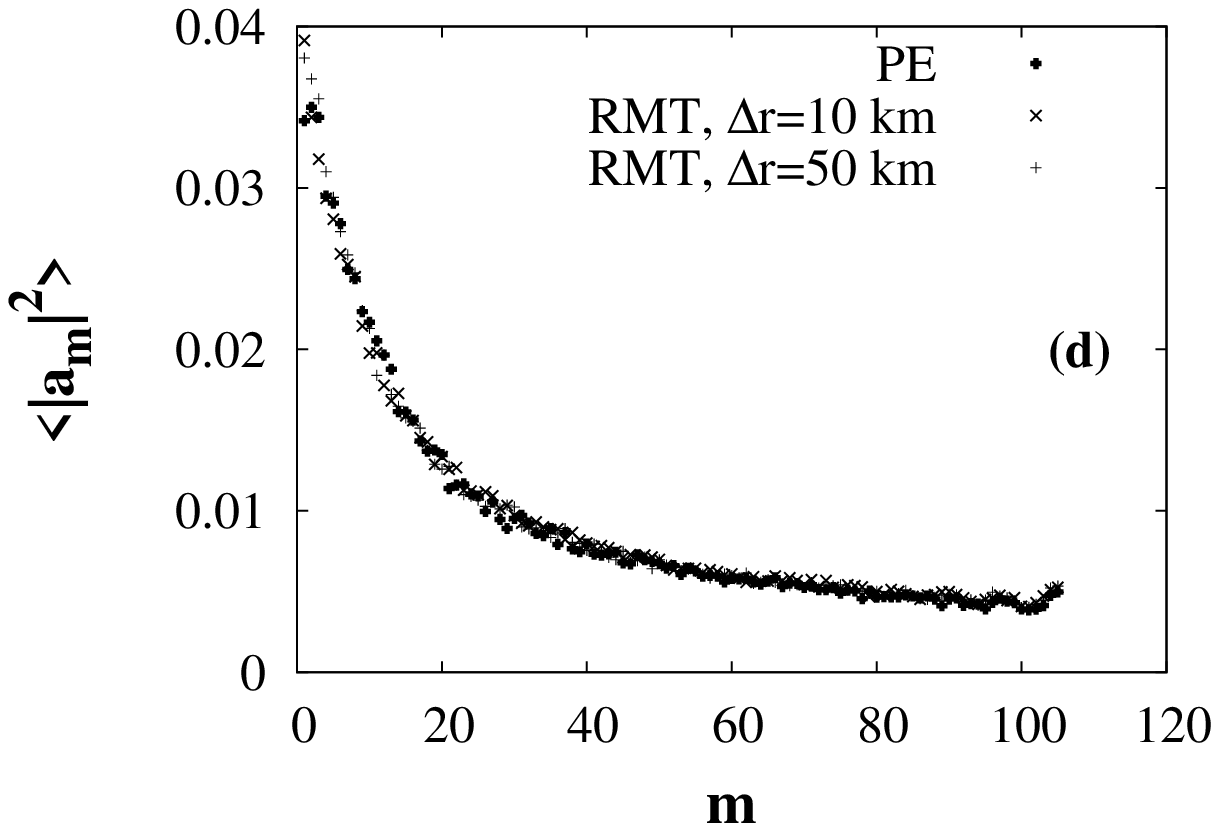}
\caption{Distribution of acoustic energy over modes at the range of 1000 km. 
Panels (a) and (b) depict results for the BEP model,
panels (c) and (d) correspond to the Munk model.
Signal frequency: 25 Hz (panels (a) and (c)) and
75 Hz (panels (b) and (d)).
The curves obtained via
the direct solving of the parabolic equation are denoted as ``PE''.
}%
\label{fig-ME}
\end{figure}

Figure \ref{fig-1mode} demonstrates range dependences of the ensemble-averaged 
first-mode energy for various waveguide models and frequencies.
The calculations were performed using Eq.~(\ref{amod})
for the initial state corresponding to the point source located at the waveguide axis, i.~e.
at the depth of 1000 meters.
For all cases, the first-mode energy decays with range that reveals tendency towards the equipartition regime \cite{Dozier_Tappert}.
In the case of 25 Hz the rate of decay is lower indicating weaker scattering, than in the case of 75 Hz.
Notably, direct Monte-Carlo simulations expose a little bit faster decay
as compared with modelling via the random matrix theory. Thus, one can conclude that the random matrix theory slightly underestimates scattering.
However, distributions of mode energies over modes show that the differences are mainly concerned with the lowest modes, 
while higher ones have nearly the same weights for all methods of simulation (see Fig.~\ref{fig-ME}).

\section{Propagator spectrum}
\label{Spectrum}

Given a realization of random inhomogeneity,
properties of sound propagation are reflected
in the spectrum of the propagator $\hat G$.
The corresponding eigenvalues and eigenfunctions obey the equation
\begin{equation}
\hat G(r_0,r\F)\Phi_n(z)=g_n(r_0,r\F)\Phi_n(z).
\label{eigen}
\end{equation}
Taking into account the mathematical equivalence between the parabolic equation (\ref{parabolic}) and
the Schr\"odinger equation, we can study spectral properties of the propagator by means of the methods developed in quantum mechanics.
In particular, we can utilize the spectral theory of quantum chaos that is mainly based on the theory of random matrices.

The eigenvalues can be found as eigenvalues of the propagator matrix $\mathbf{G}$.
Owing to the unitarity, they can be recast as
\begin{equation}
 g_n=e^{-i\varphi_n}, \quad 
\varphi_n\in\Re.
\label{fn}
\end{equation}
Since eigenvalues of the propagator belong to the unit circle in the complex plane,
its matrix corresponds to the circular ensemble of random matrices \cite{Stockman}.

It is important to note the equivalence between spectrum of $\hat G(0,r\F)$ and spectrum of 
the Floquet operator governing wave evolution in a fictitious range-periodic waveguide with sound-speed inhomogeneity
described as
\begin{equation}
 \bar \delta c(r'+jr\F,z) = \delta c(r',z),\quad
 0\le r'\le r\F, \quad j=1,2,3\text{...},
 \label{fiction}
\end{equation}
where $\delta c(r,z)$ corresponds to some individual realization of the actual (i.~e. not range-periodic) waveguide.
This property allows for interpretation of propagator spectrum in terms of the well-developed Floquet theory and thereby
significantly facilitates the analysis of coherence in wave dynamics.
Floquet theory was utilized for studying wave chaos in sound propagation in the ocean in Refs.~\cite{Viro05,PRE76,SibFU}.
It has to be kept in mind that this equivalence holds only if we restrict ourselves by the range interval $r\in[0:r\F]$.
However, the final range $r\F$ of the propagator can be chosen arbitrarily.
So, if we want to study long-range wave evolution, then we have to set a corresponding value of $r\F$.

As it was argued in Refs.~\cite{UFN,PRE87},
 wave chaos associated with scattering on internal waves
 reveals itself
 in the statistics of level spacings.
A level spacing for the circular ensemble is defined as
\begin{equation}
\begin{gathered}
 s=\frac{k_0M(\varphi_{m+1}-\varphi_m)}{2\pi},\quad 
m = 1,2,\dots,M, \\
\varphi_{M+1} = \varphi_1 + \frac{2\pi}{k_0}.
\end{gathered}
\label{spacing}
\end{equation}
where the sequence of eigenphases $\varphi_m$ is rearranged  in the ascending order,
$M$ is the total number of eigenvalues for a single realization of the propagator, equal to the number
of trapped modes.
Statistical distribution of level spacings is connected to all $m$-order correlation functions \cite{Stockman}
\begin{displaymath}
 R(\varphi_1,\text{...},\varphi_m) = \frac{M\text{!}}{(M-m)\text{!}}\int P(\varphi_1,\text{...},\varphi_M)\,d\varphi_{m+1},\text{...},d\varphi_M,
\end{displaymath}
where $P(\varphi_1,\text{...},\varphi_m)$ is the joint eigenphase probability distribution.
Hence, level spacing statistics serves as a good indicator of differences between the spectrum of the propagator constructed via the RMT
and the actual propagator obtained by solving the parabolic equation.
Furthermore, level spacing distribution qualitatively depends on the strength of mode coupling.
For instance, if scattering on inhomogeneity
is weak and modes don't significantly interact, the corresponding eigenphases of the propagator are statistically 
independent from each other, and level spacing distribution obeys the Poisson law
\begin{equation}
  \rho(s)\sim\exp(-s).
\label{Poisson}
\end{equation}
In the opposite case of strong scattering and global cross-mode coupling, the neighboring eigenphases ``repulse'' that leads 
to level spacing statistics described by the Wigner surmise
\begin{equation}
\rho(s)\sim s^{\alpha}\exp\left(-Cs^2\right),
\label{surmise}
\end{equation}
where constants $\alpha$ and $C$ depend on symmetries of the propagator.
$\rho(s)$ is normalized as 
\begin{equation}
 \int \rho(s)ds=1,\quad
 \int s\rho(s)ds=1.
\end{equation}
As the unitarity is the only constraint on the propagator, 
the propagator corresponds to the circular unitary ensemble (CUE), 
and we can set $\alpha=2$ and $C = 4/\pi$ \cite{Kol97}.
 
If the matrix of the propagator can be divided into two blocks, one corresponding to modes experiencing strong scattering,
 and near-diagonal one corresponding to modes maintaining coherence, then
the resulting level spacing distribution has some intermediate form between the Poisson and Wigner statistics.
In the high-frequency case one can use the Berry-Robnik distribution \cite{BR}
\begin{equation}
\rho(s)= \left[
v_{\mathrm{r}}^2\operatorname{erfc}\left(\frac{\sqrt{\pi}}{2}v_{\mathrm{c}}s\right)+\left(2v_{\mathrm{r}}v_{\mathrm{c}}+\frac{\pi}{2}v_{\mathrm{c}}^3s\right)
\exp\left(-\frac{\pi}{4}v_{\mathrm{c}}^2s^2\right)
\vphantom{\frac{\sqrt{\pi}}{2}}\right]\exp(-v_{\mathrm{r}}s),
\label{berrob}
\end{equation}
where $v_\mathrm{c}$ and $v_\mathrm{r}$ denote fractions of strongly and weakly scattered modes, respectively, $v_{\mathrm{r}}+v_{\mathrm{c}}=1$.
If there is good agreement between ray and wave descriptions, 
$v_\mathrm{r}$ is equal to relative volume of regular dynamics in phase space of ray equations,
while $v_\mathrm{c}$ is equal to relative phase space volume attributed to chaotic ray dynamics.
Berry-Robnik distribution undergoes the smooth transition
from the Poisson to the Wigner law as $v_{\mathrm{r}}$ decreases from 1 to 0.
Strictly speaking, the case of $v_{\mathrm{r}}=0$ corresponds to the Wigner distribution with $\alpha=1$, that 
is not identical but roughly similar to the case of $\alpha=2$. 
Thus, we can track scattering-induced decoherence of a wavefield by fitting level spacing distribution
using the Berry-Robnik formula (\ref{berrob}).
It is worth reminding that long-range sound propagation is commonly realized with low-frequency sound, when 
one shouldn't expect excellent correspondence between wave motion and the underlying ray dynamics.
Therefore, the formula (\ref{berrob}) allows us to find only approximate fraction 
of weakly scattered modes.

Another way to track the transition from the Poisson to Wigner statistics is the approximation of level spacing distribution
using the Brody formula
\begin{equation}
\rho(s)=(\beta+1)A_{\beta}s^{\beta}\exp(-A_{\beta}s^{\beta+1}),
 \label{brody}
\end{equation}
where
\begin{equation}
A_{\beta}=\left[\Gamma\left(\frac{\beta+2}{\beta+1}\right)\right]^{\beta+1},
\label{abeta}
\end{equation}
$\Gamma(\mathrm{...})$ is the Euler gamma function. 
Eq.~(\ref{brody}) undergoes
the transition from the Poisson to the Wigner law with increasing of the parameter $\beta$ from 0 to 1. 
Unfortunately, the Brody distribution is empirical, and the parameter $\beta$ doesn't have certain physical
interpretation. However, formula (\ref{brody}) often provides more accurate fit of level spacing statistics 
than the Berry-Robnik and other formulae \cite{Stockman,Prosen}.
So, if we are rather interested in comparison of level spacing statistics obtained using different methods,
the Brody distribution is a proper choice for the fitting function.

Analysis of propagator eigenfunctions can be carried out without any auxiliary assumptions and approximations, in contrast
to analysis of eigenvalues. Therefore, it can be considered as a more robust tool.
Eigenfunctions of the propagator can be expanded over normal modes of the background waveguide
\begin{equation}
 \Phi_n=\sum\limits_m C_{mn}\psi_m,
 \label{eigf}
\end{equation}
where $C_{mn}$ is $m$-th element of the $n$-th eigenvector of the propagator matrix.
Introduce a quantity 
\begin{equation}
\mu=\sum\limits_{m=1}^M\lvert C_{mn}\rvert^2m,
 \label{mu}
\end{equation}
that can serve as an identifier of a propagator eigenfunction \cite{Viro05}.
In an unperturbed waveguide,
only one normal mode contributes to each eigenfunction, and
$\mu$ coincides with the number of this mode.

According to the definition (\ref{eigen}), any eigenfunction of $\hat G$ remain invariant after propagation from $r=0$
to $r=r\F$. Its form qualitatively depends on scattering on inhomogeneity.
As scattering determines rate of mode coupling,
an eigenfunction corresponding to propagation with strong impact of scattering 
is compound of many normal modes of the background waveguide.
Hence, we can quantify strength of scattering by means of the participation ratio in the expansion (\ref{eigf}).
Participation ratio of the $n$-th eigenfunction is calculated as
\begin{equation}
\nu = \left(
\sum\limits_{m=1}^M\lvert C_{mn}\rvert^4
\right)^{-1}.
 \label{npc}
\end{equation}
$\nu$ is equal to 1 in a range-independent waveguide, and increases
as scattering intensifies.

Statistical analysis of the propagator's spectrum for a model of the underwater sound channel
in the Sea of Japan was performed in Ref.~\cite{PRE87}.
It was shown that spectral statistics qualitatively changes with increasing $r\F$ indicating onset of
global incoherent mode coupling, excepting for a small group of low-number modes associated with a weakly
divergent beam.

\section{RMT modeling vs direct calculation. II. Spectral statistics}
\label{Numer}

The present section represents comparative analysis of the propagator spectral characteristics obtained
via the direct calculation and those obtained by means of the random matrix theory. 
Level spacing distributions were averaged in the following way:
\begin{equation}
\rho(s)=\lim\limits_{N\to\infty}\frac{1}{N}\sum\limits_{n=1}^{N} \rho_n(s),
\label{aver_ps}
\end{equation}
where $\rho_n(s)$ is a level spacing distribution
corresponding to the $n$-th realization of the propagator, $N=1000$ is number of realizations.

\subsection{Signal frequency 25 Hz}

\begin{figure}[!htb]
\includegraphics[width=0.32\textwidth]{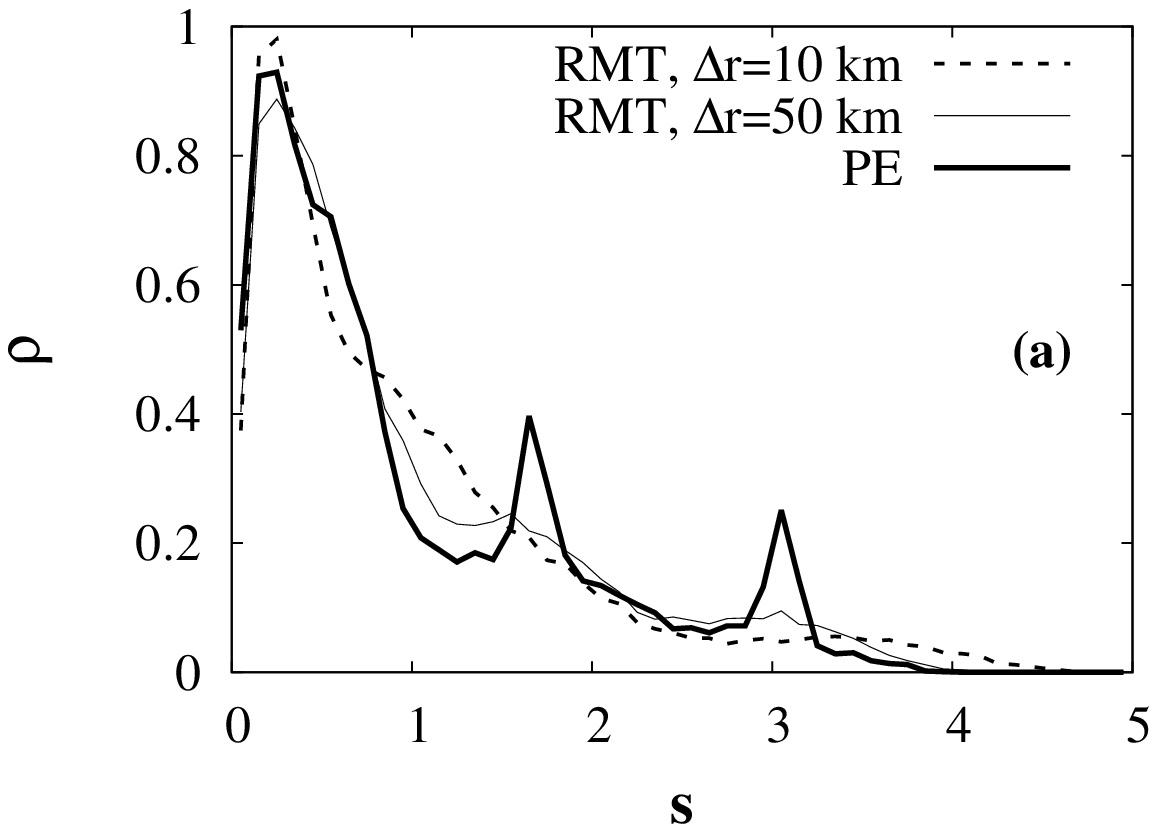}
\includegraphics[width=0.32\textwidth]{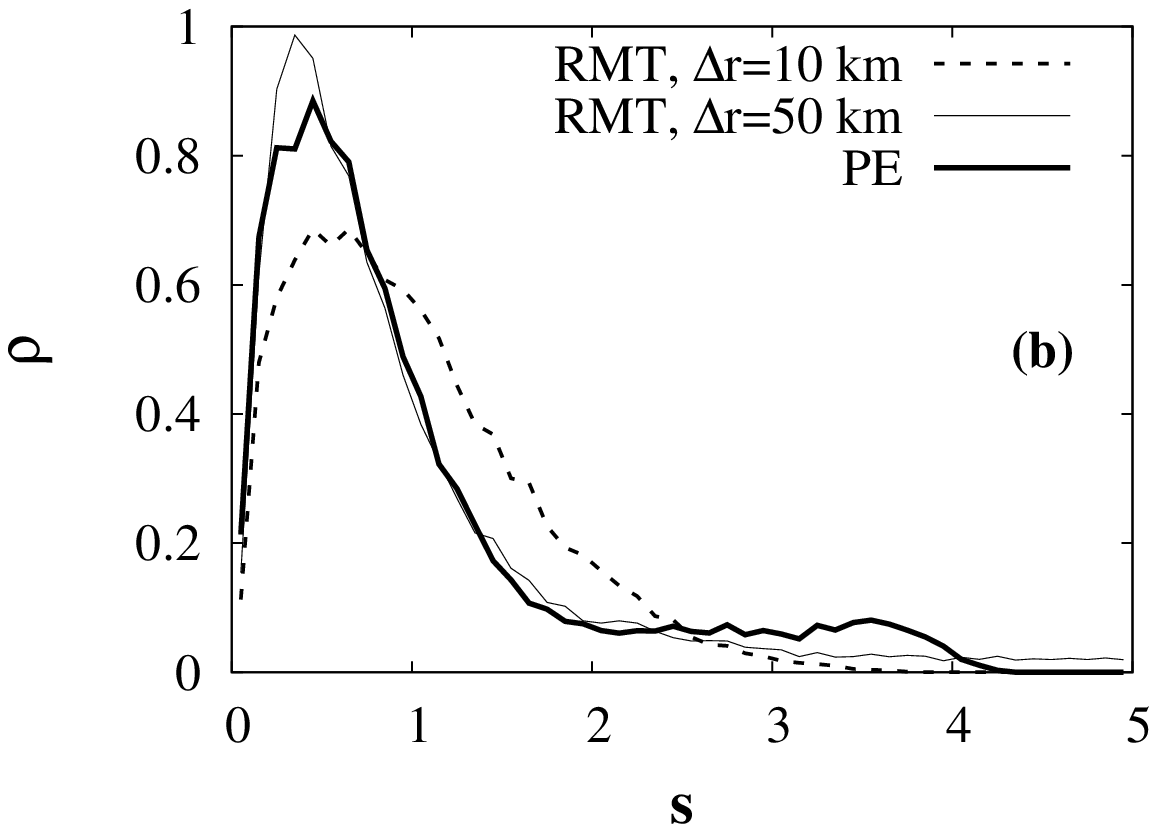}
\includegraphics[width=0.32\textwidth]{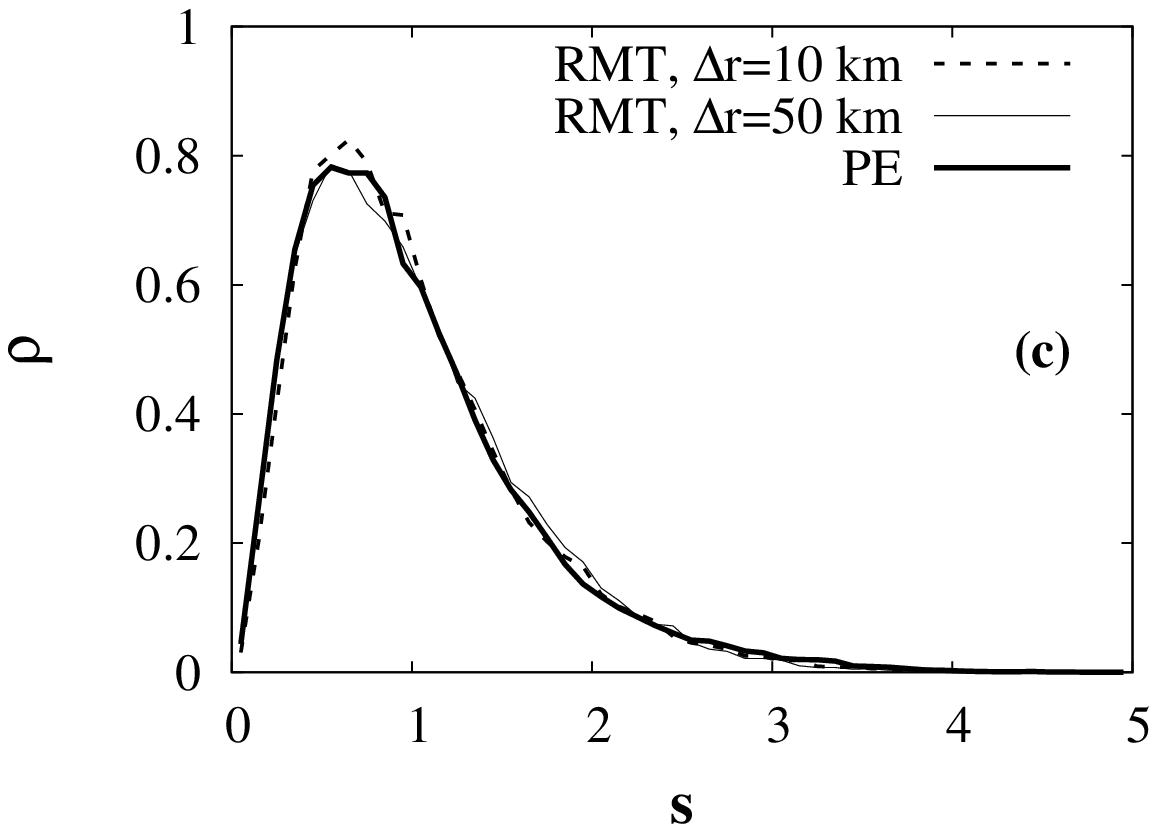}\\
\includegraphics[width=0.32\textwidth]{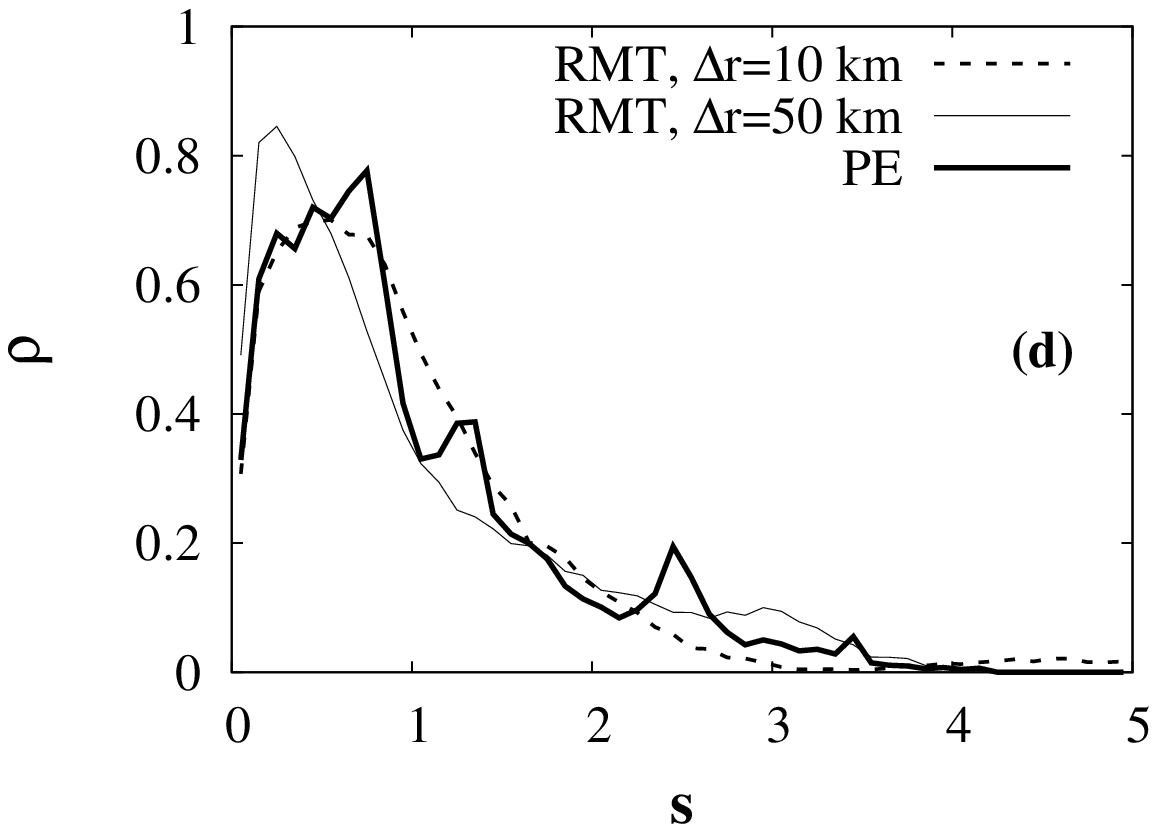}
\includegraphics[width=0.32\textwidth]{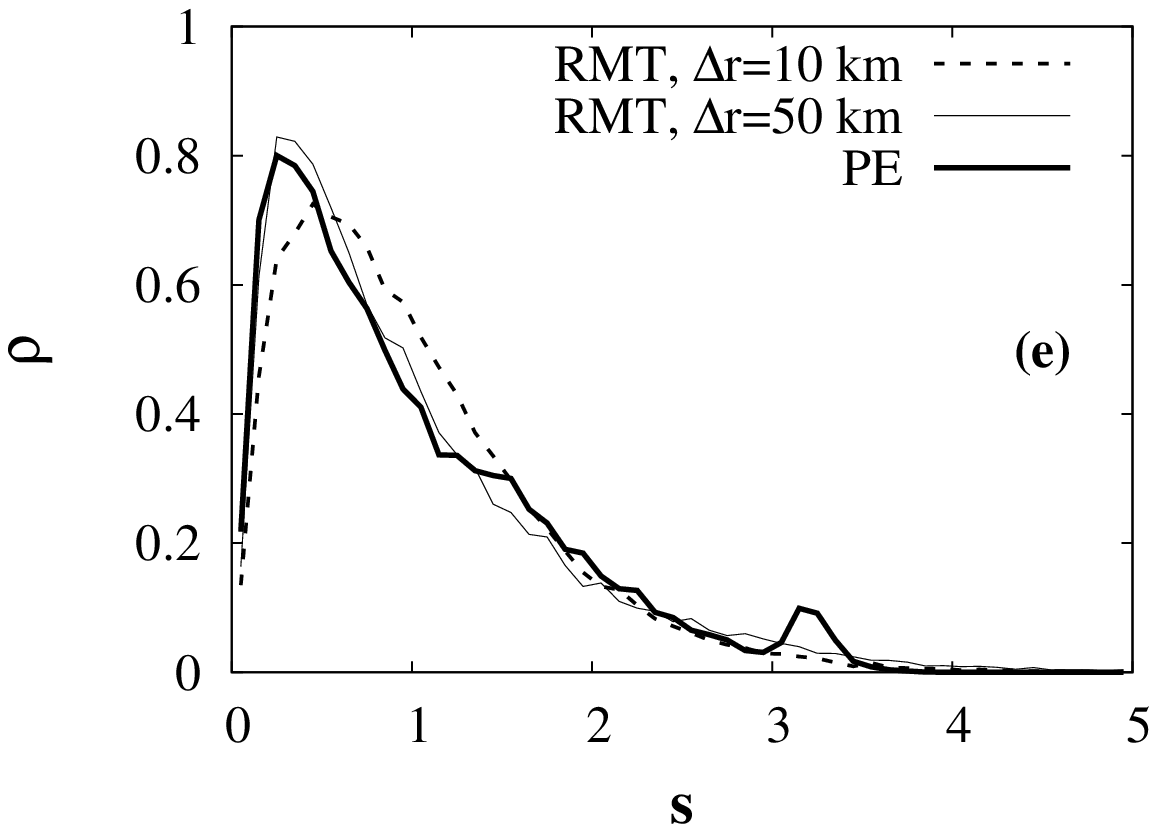}
\includegraphics[width=0.32\textwidth]{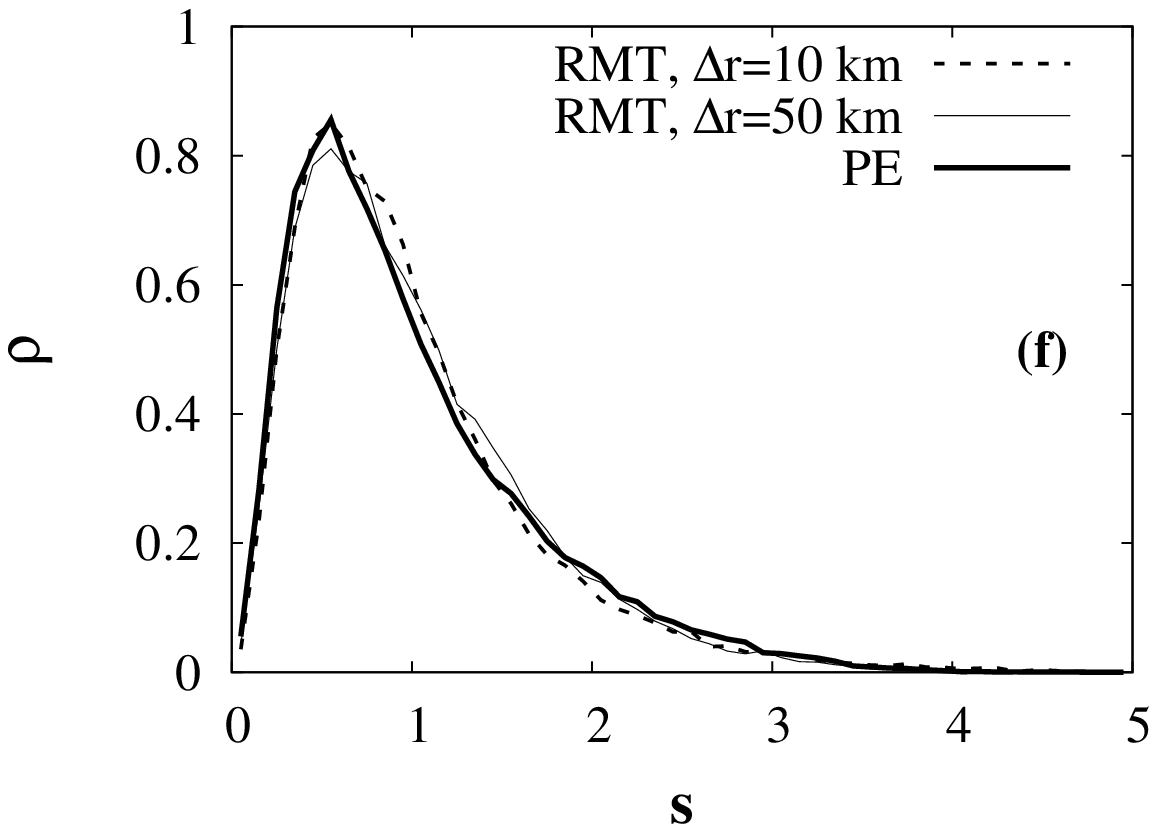}

\caption{Level spacing distribution for the BEP (upper raw) and Munk (lower raw) models.
Signal frequency is of 25 Hz. Propagation distances $r\F$: (a) and (d) $r\F=50$~km, (b) and (e) $r\F=250$~km, (c) and (f)  $r\F=1000$~km. The curves obtained via
the direct solving of the parabolic equation are denoted as ``PE''.}%
\label{fig-distr25}
\end{figure}
\begin{figure}[!htb]
\includegraphics[width=0.47\textwidth]{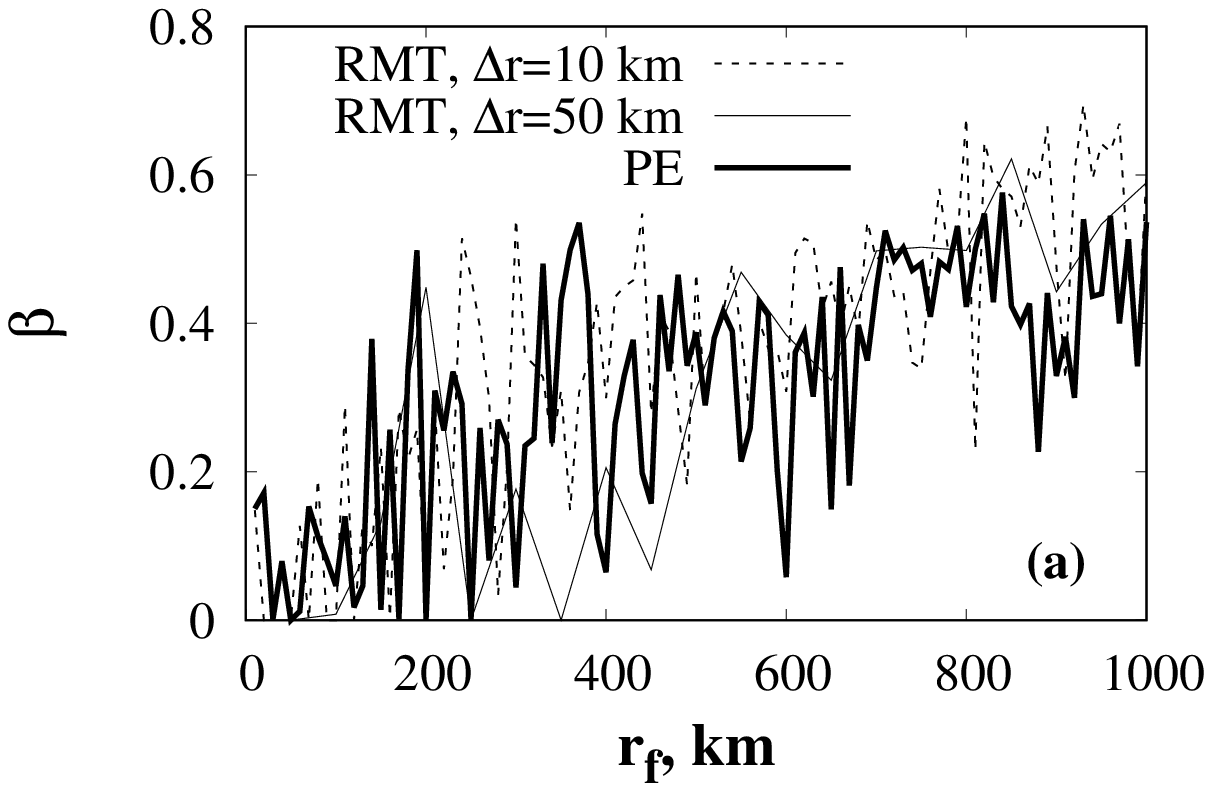}
\includegraphics[width=0.47\textwidth]{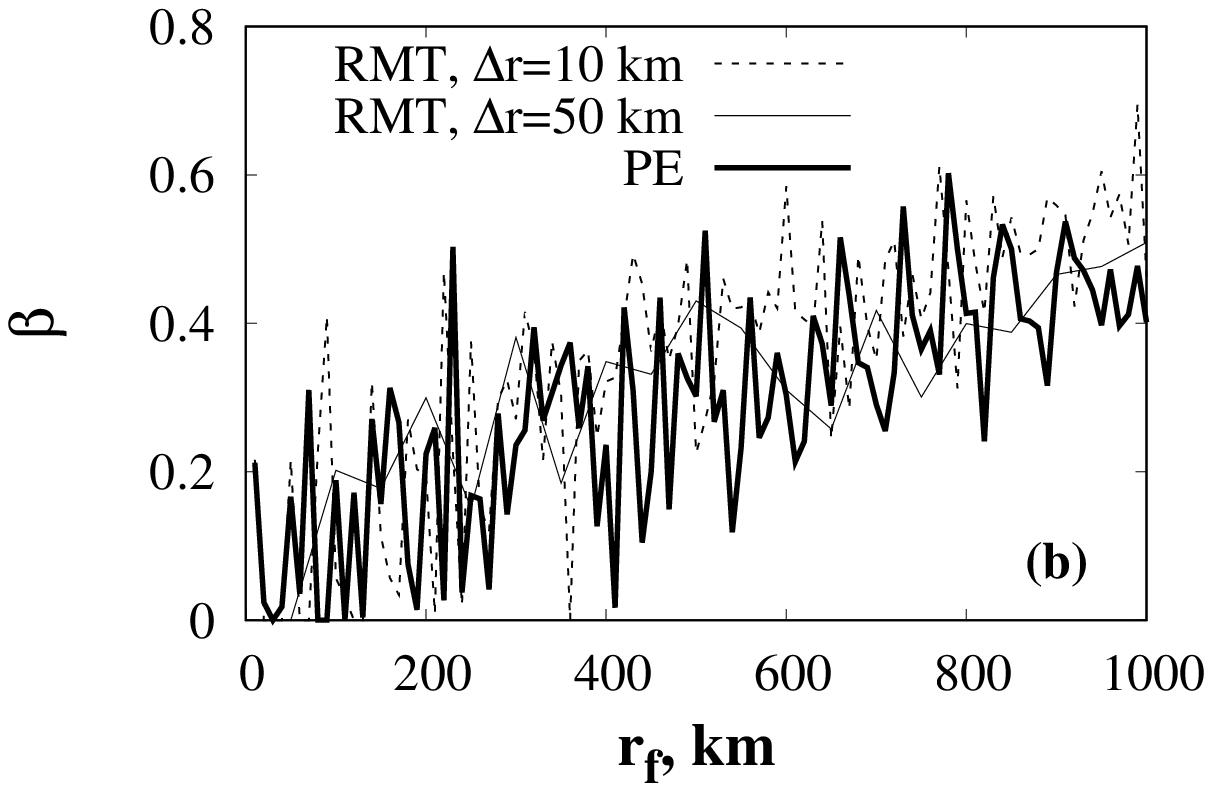}
\caption{Range dependence of the Brody parameter $\beta$ corresponding to the best fit of level spacing distribution.
Panel (a) corresponds to the BEP model, panel (b) depicts results for the canonical Munk waveguide. Signal frequency is of 25 Hz.
The curves obtained via the direct solving of the parabolic equation are denoted as ``PE''.}%
\label{fig-sp25}
\end{figure}

Figure \ref{fig-distr25} demonstrates level spacing distributions for various ranges. 
Markedly, there is discrepancy between the results of direct simulation and RMT modeling
for short ranges. In particular,
direct simulation reveals 
peaks for relatively large spacing values (approximately 1.7 and 3.0) that are 
absent in the curves obtained with RMT. 
The discrepancy ceases with increasing range, and finally, for $r\F=1000$~km,
the distributions almost completely coincide. The distributions for $r\F=1000$~km reveal substantial eigenvalue repulsion,
as in the case of the Wigner surmise (\ref{surmise}).
However, fitting of the distributions by means of the Brody formula (\ref{brody}) results in strongly fluctuating
range dependence of the Brody parameter $\beta$ (see Fig.~\ref{fig-sp25}). Almost the same behavior is also observed for 
the fitting using the Berry-Robnik formula (not shown). Presence of strong fluctuations can indicate inconsistence of level spacing statistics 
with the aforementioned analytical expressions.

\begin{figure}[!htb]
\includegraphics[width=0.32\textwidth]{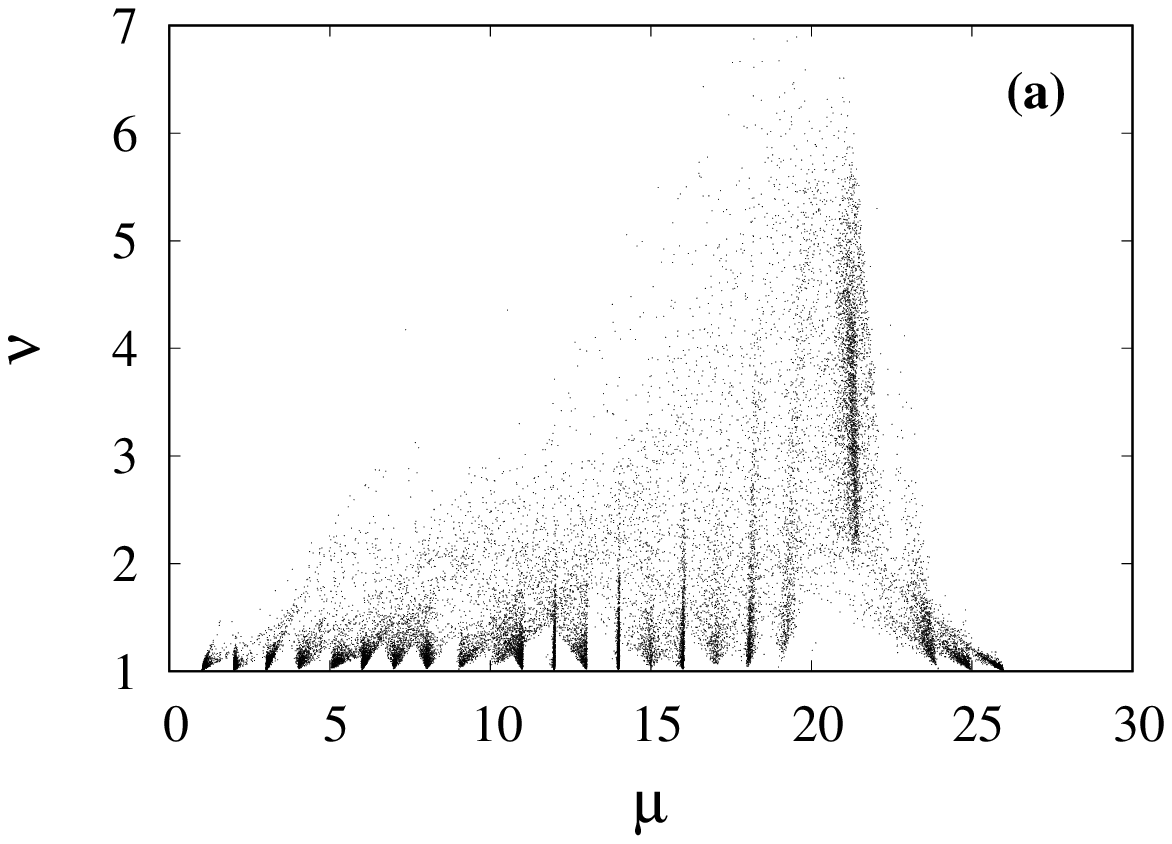}
\includegraphics[width=0.32\textwidth]{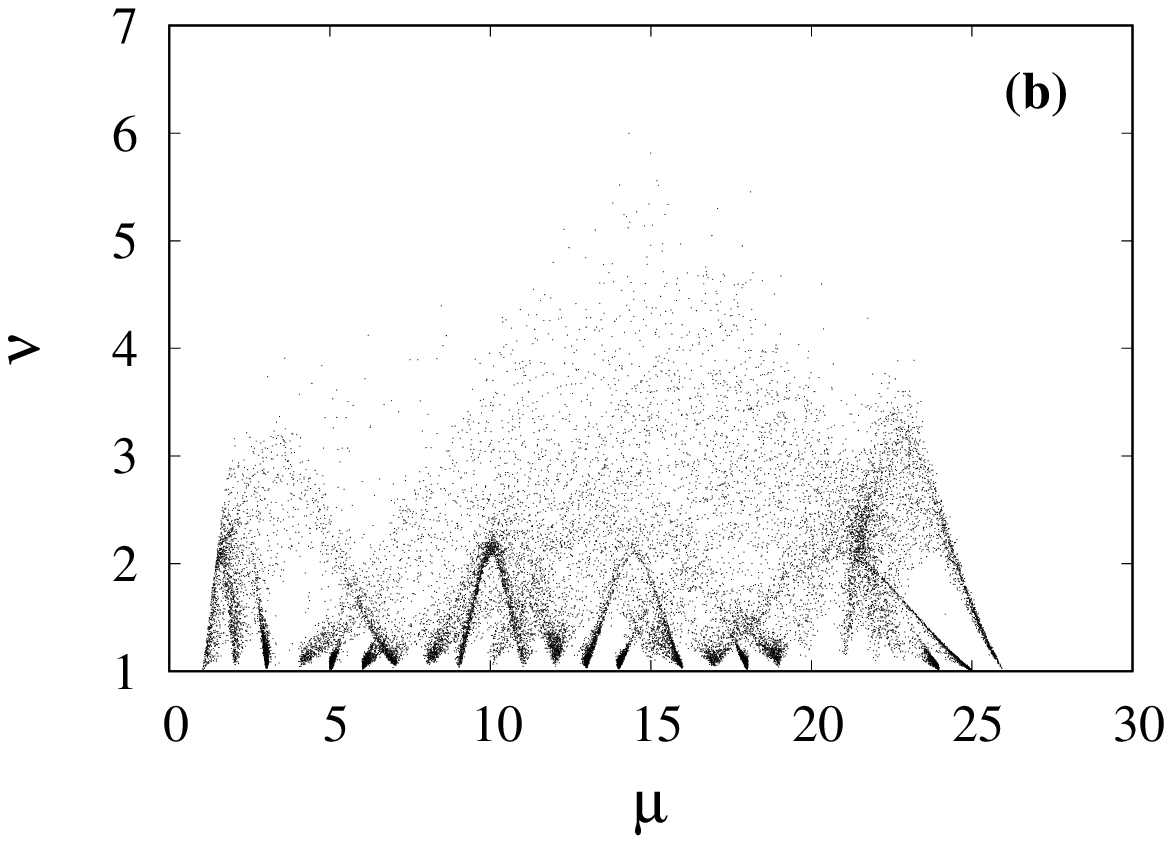}
\includegraphics[width=0.32\textwidth]{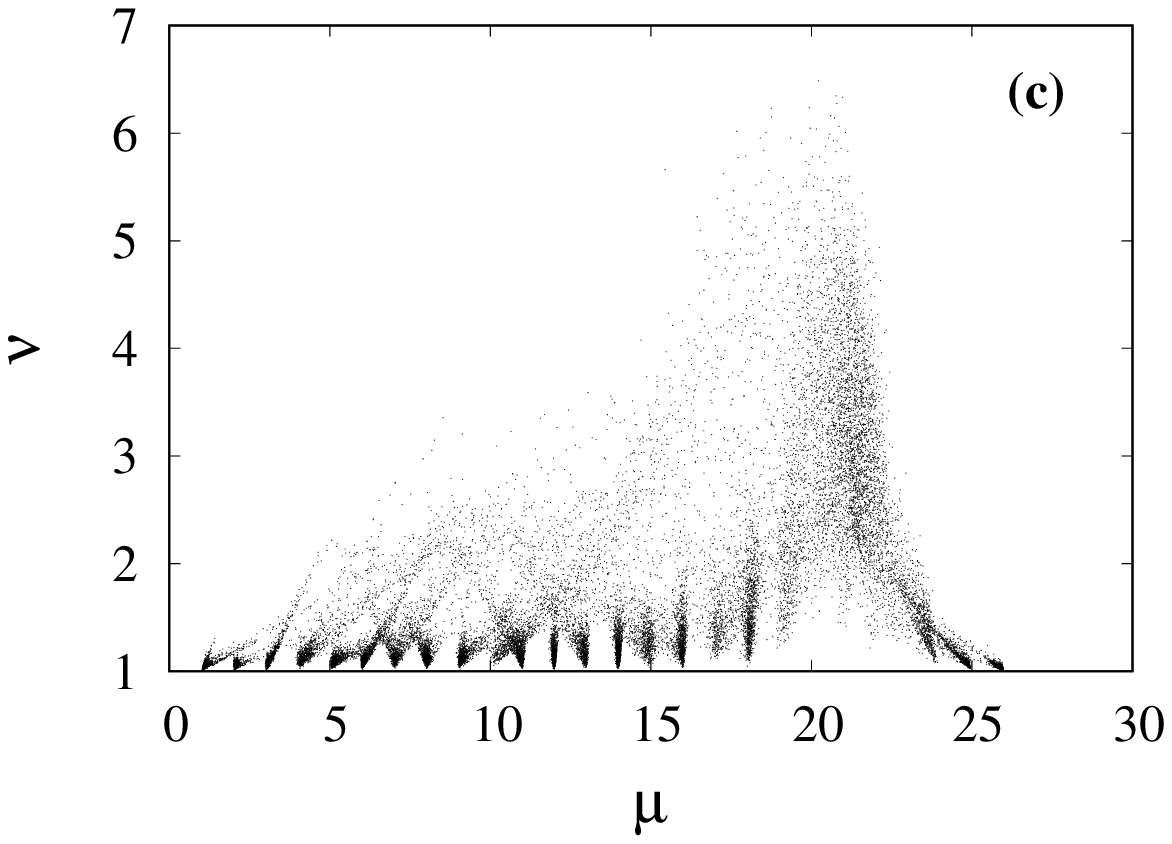}\\
\includegraphics[width=0.32\textwidth]{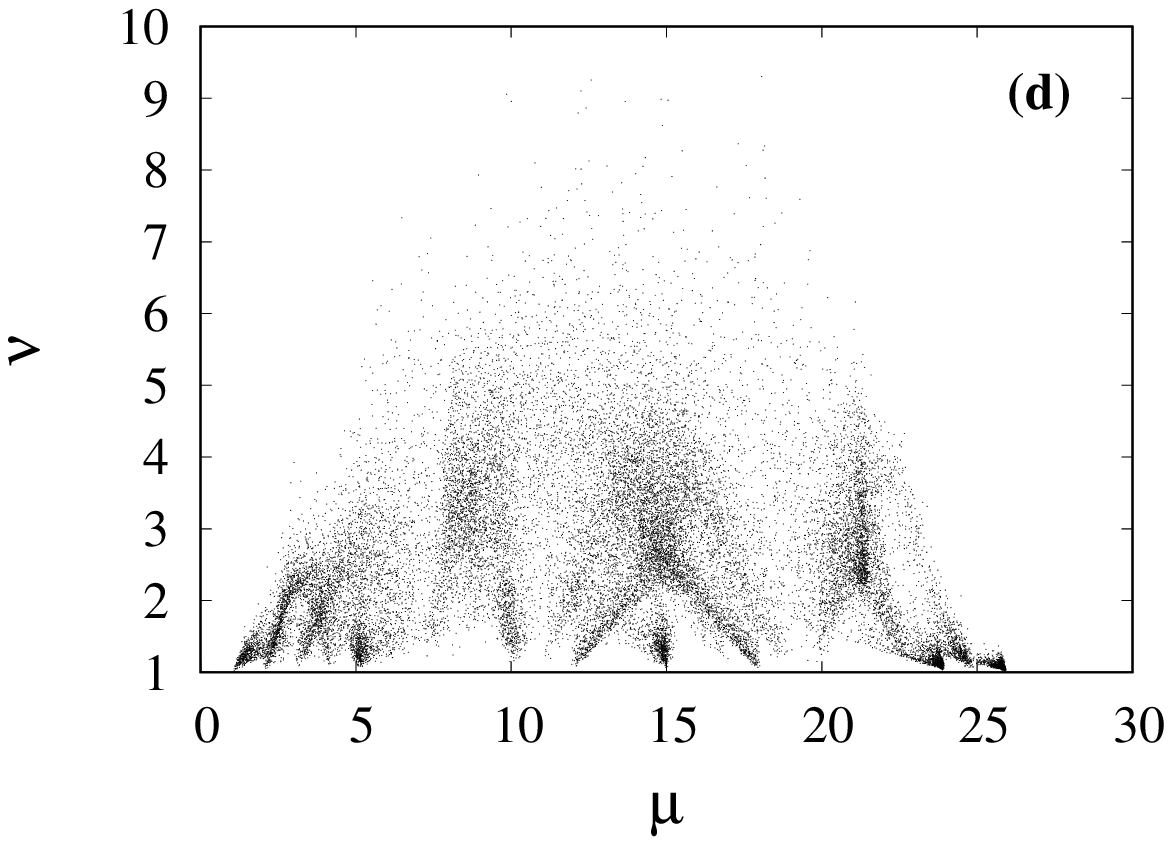}
\includegraphics[width=0.32\textwidth]{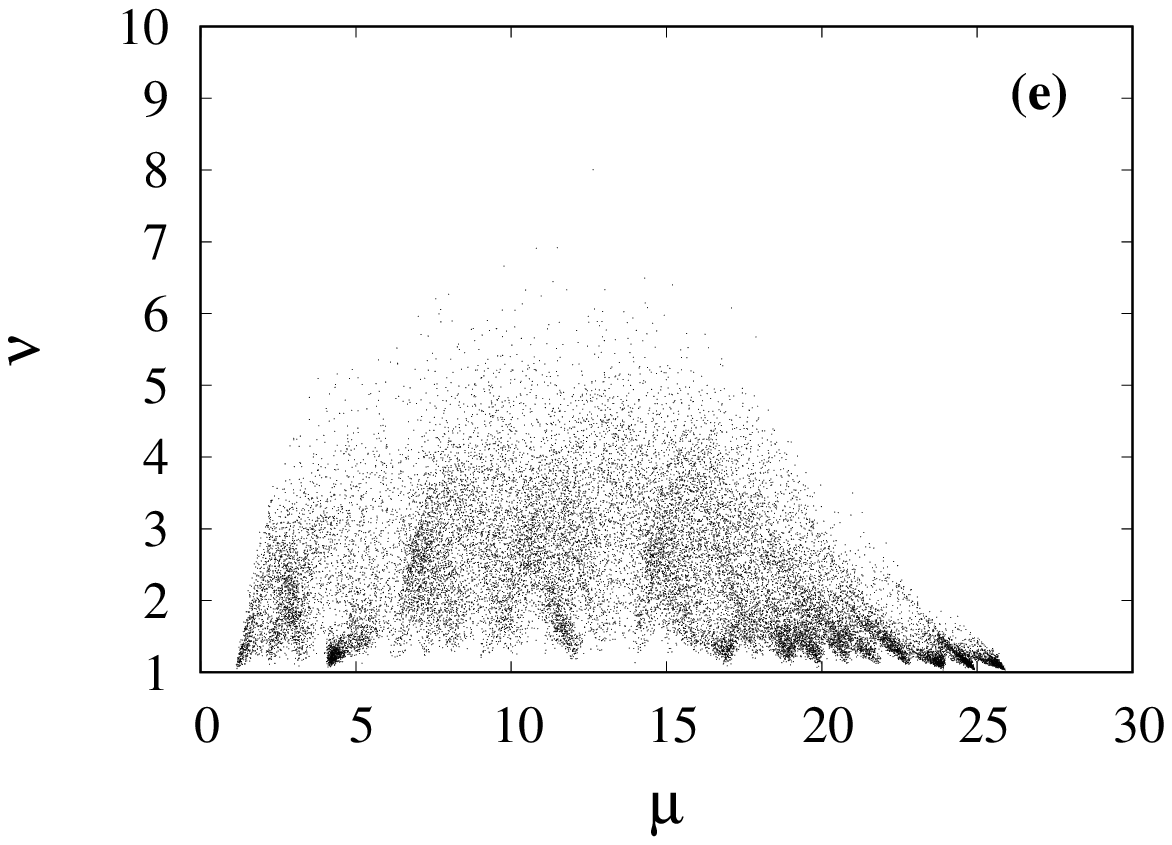}
\includegraphics[width=0.32\textwidth]{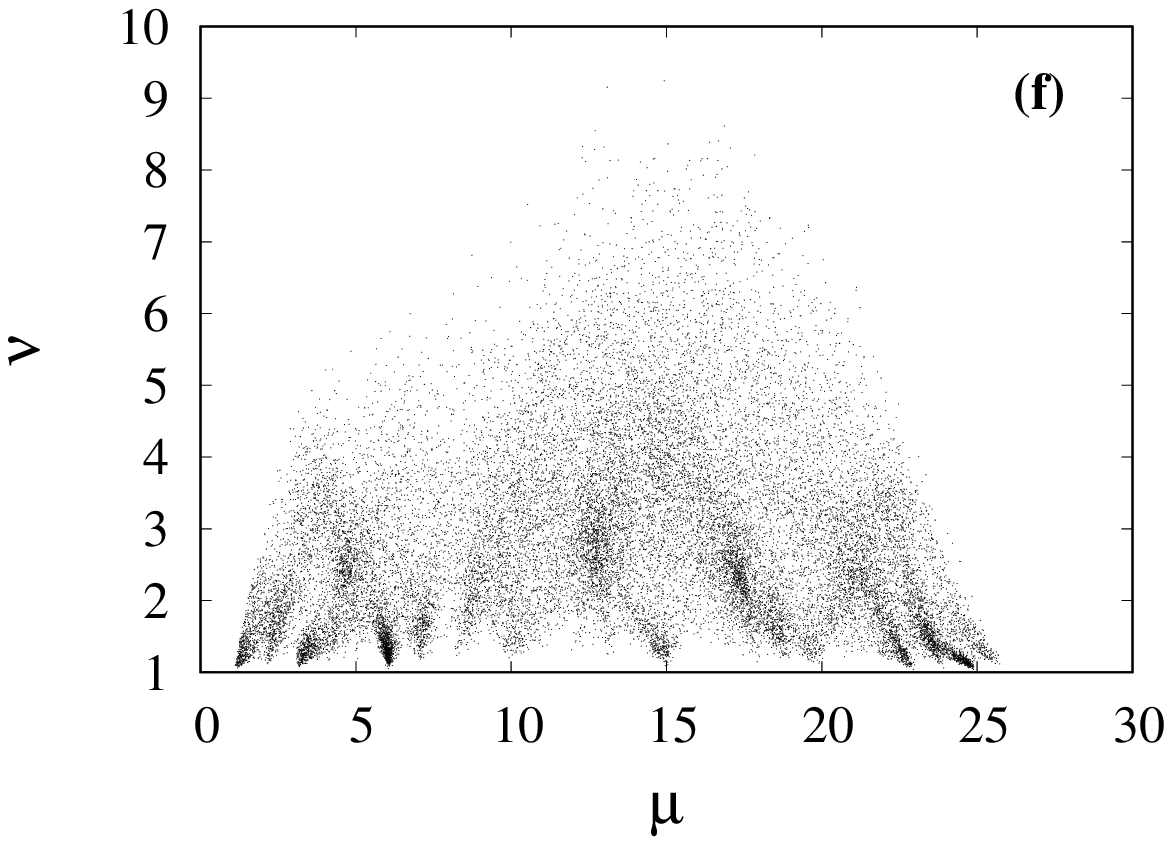}\\
\includegraphics[width=0.32\textwidth]{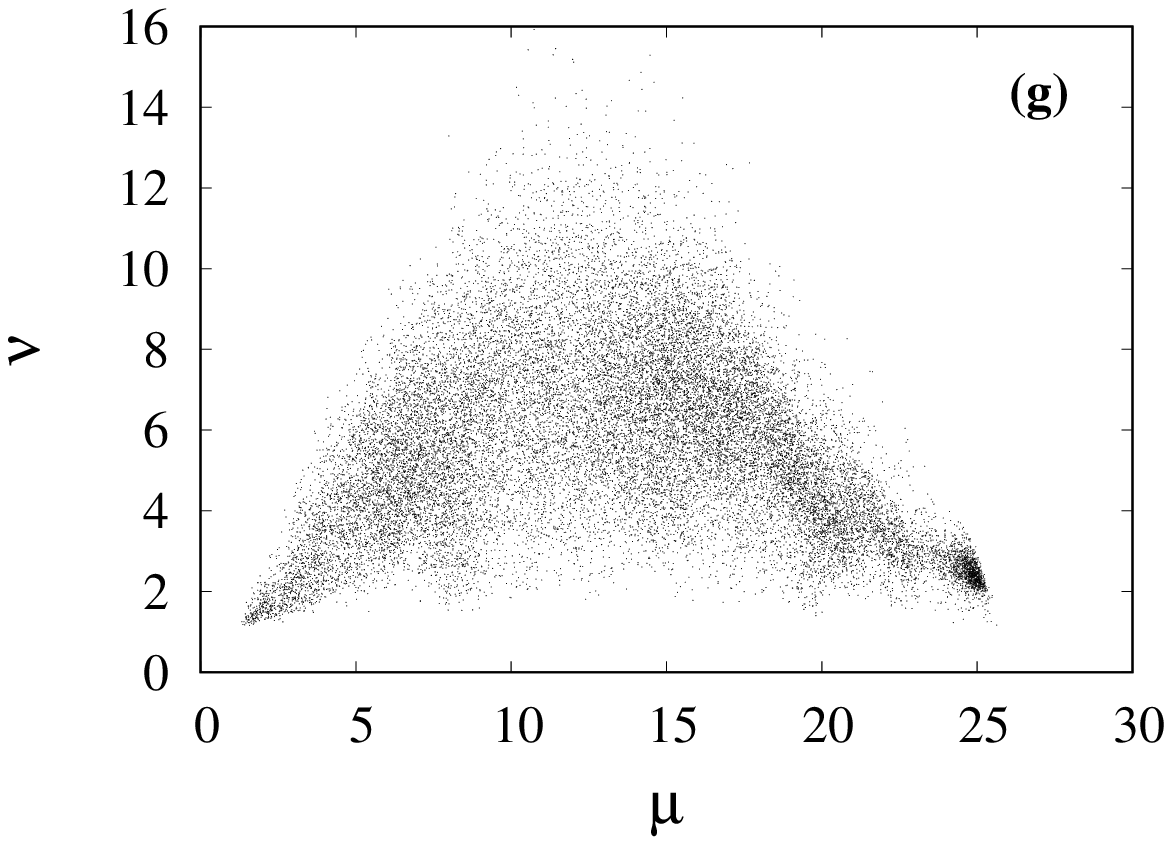}
\includegraphics[width=0.32\textwidth]{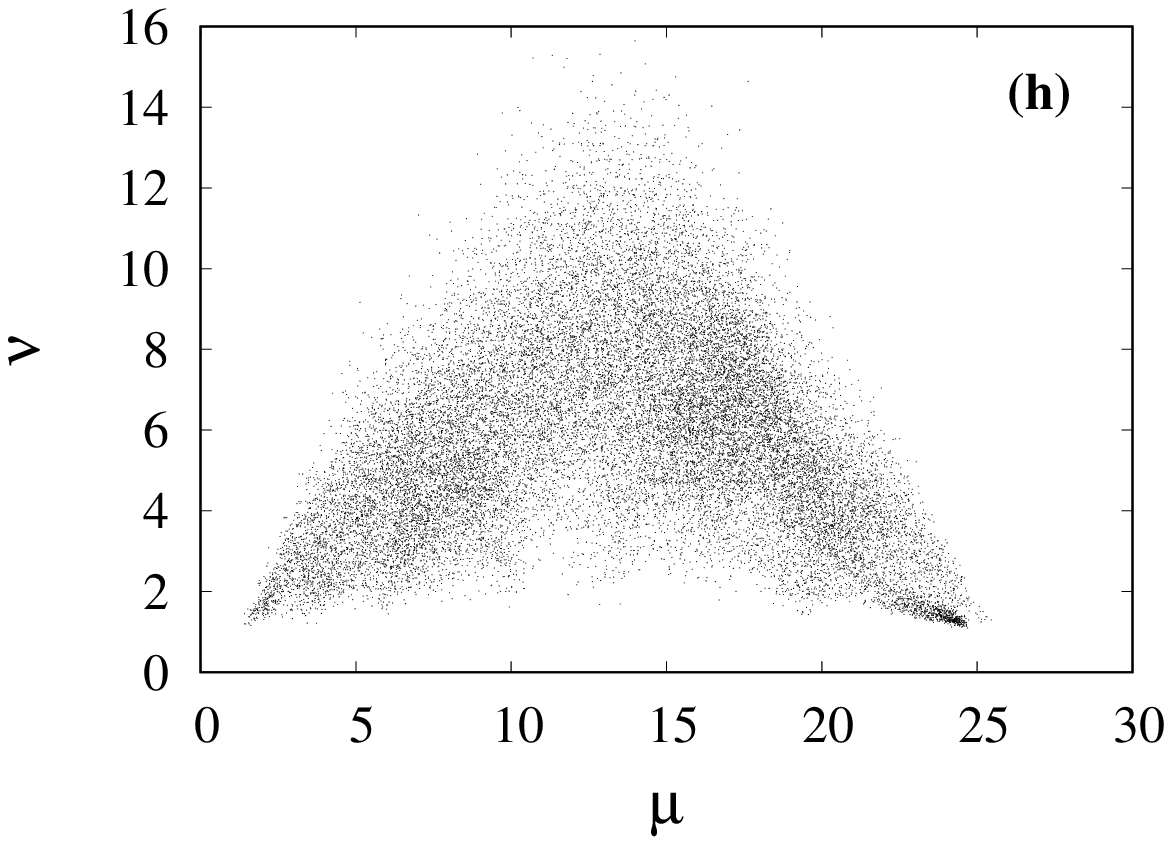}
\includegraphics[width=0.32\textwidth]{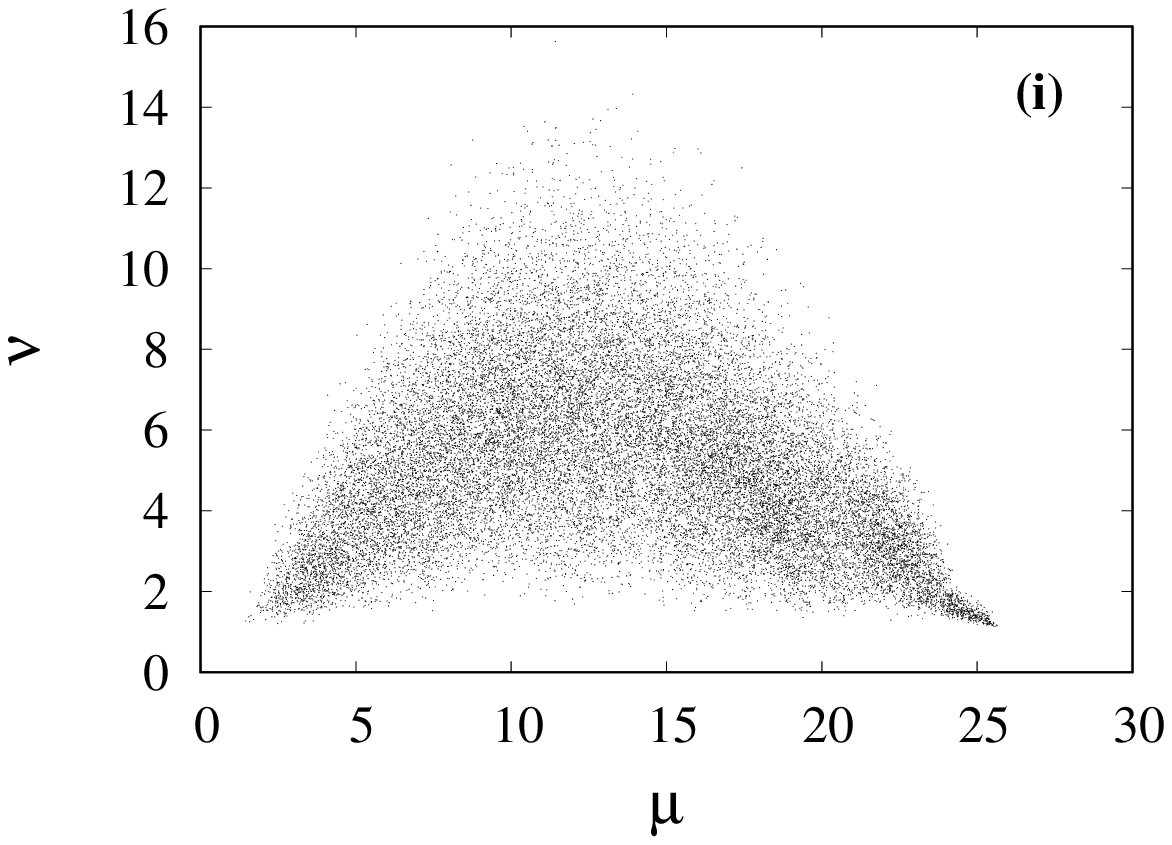}
\caption{Distribution of eigenfunctions in the $\mu$--$\nu$ plane, 
where the parameter $\mu$ is given by Eq.~(\ref{mu}), and $\nu$ is the 
participation ratio (\ref{npc}). The case of the BEC model of a waveguide.
Distance values:
(a)-(c) $r\F=50$~km, (d)-(f) $r\F=250$~km, and (g)-(i) $r\F=1000$~km.
The sound frequency is 25~Hz. Left column corresponds to numerical solution of the parabolic equation, middle and right columns correspond
to RMT modeling with $\Delta r=10$~km and $\Delta r=50$~km, respectively.}%
\label{fig-bepfunc25}
\end{figure}
\begin{figure}[!htb]
\includegraphics[width=0.32\textwidth]{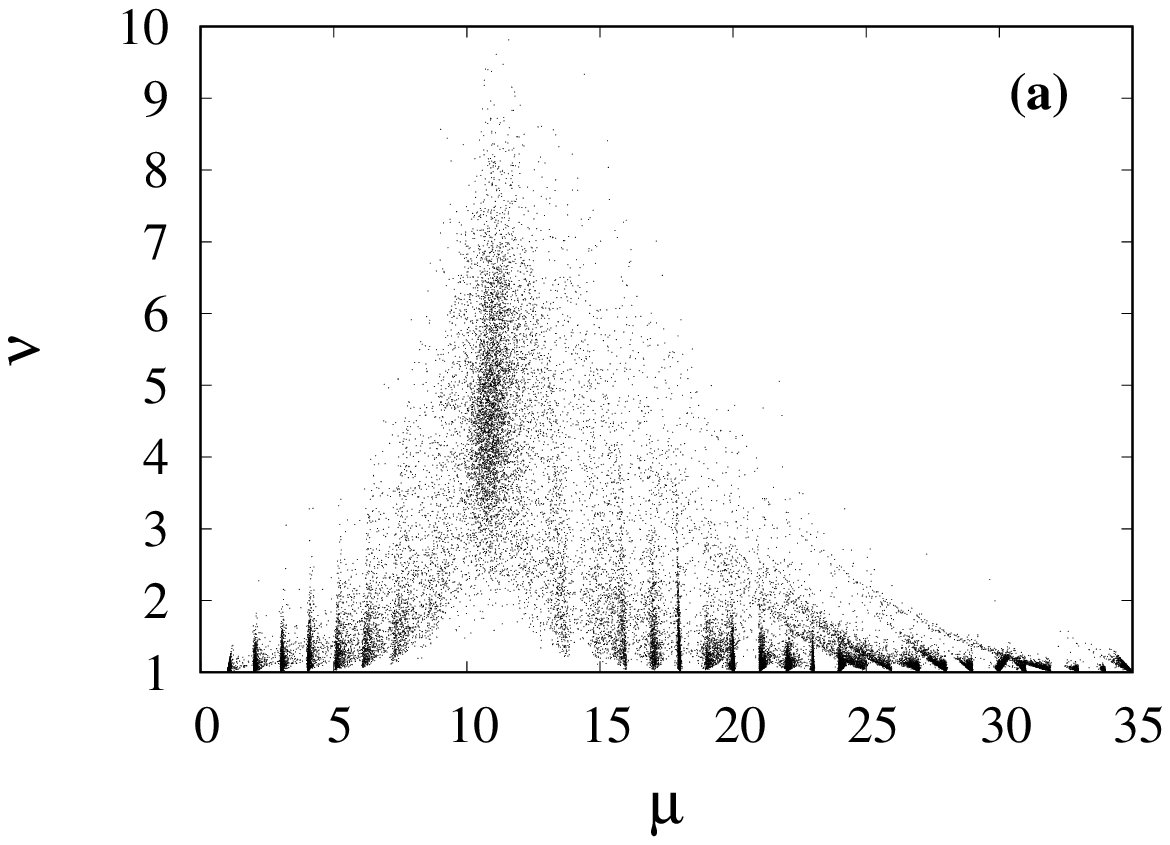}
\includegraphics[width=0.32\textwidth]{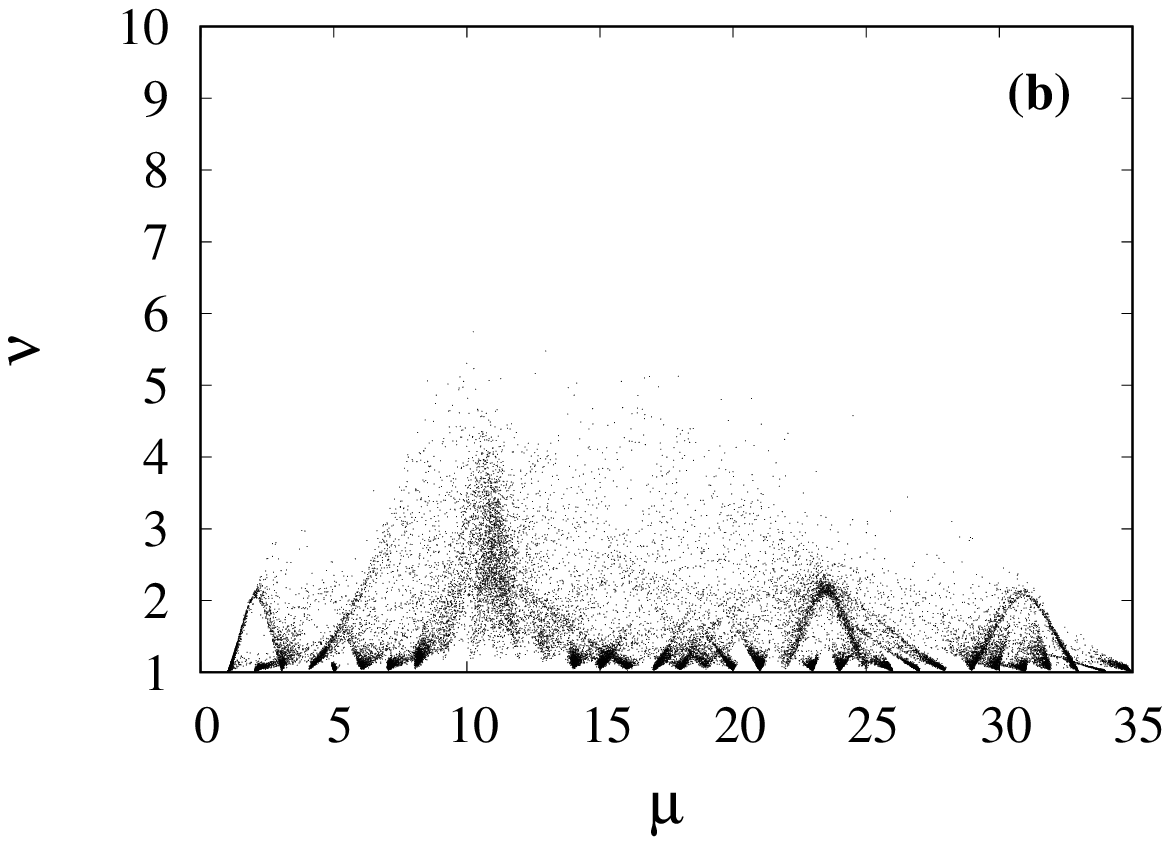}
\includegraphics[width=0.32\textwidth]{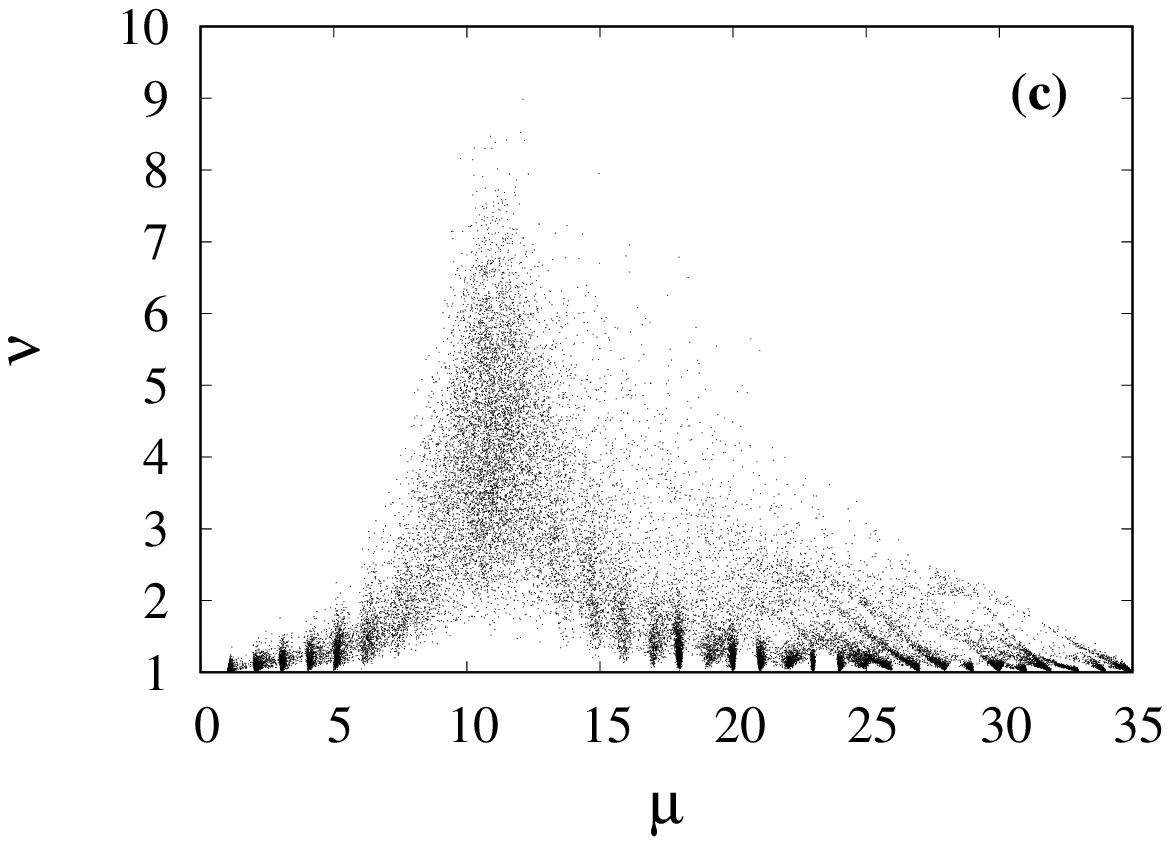}\\
\includegraphics[width=0.32\textwidth]{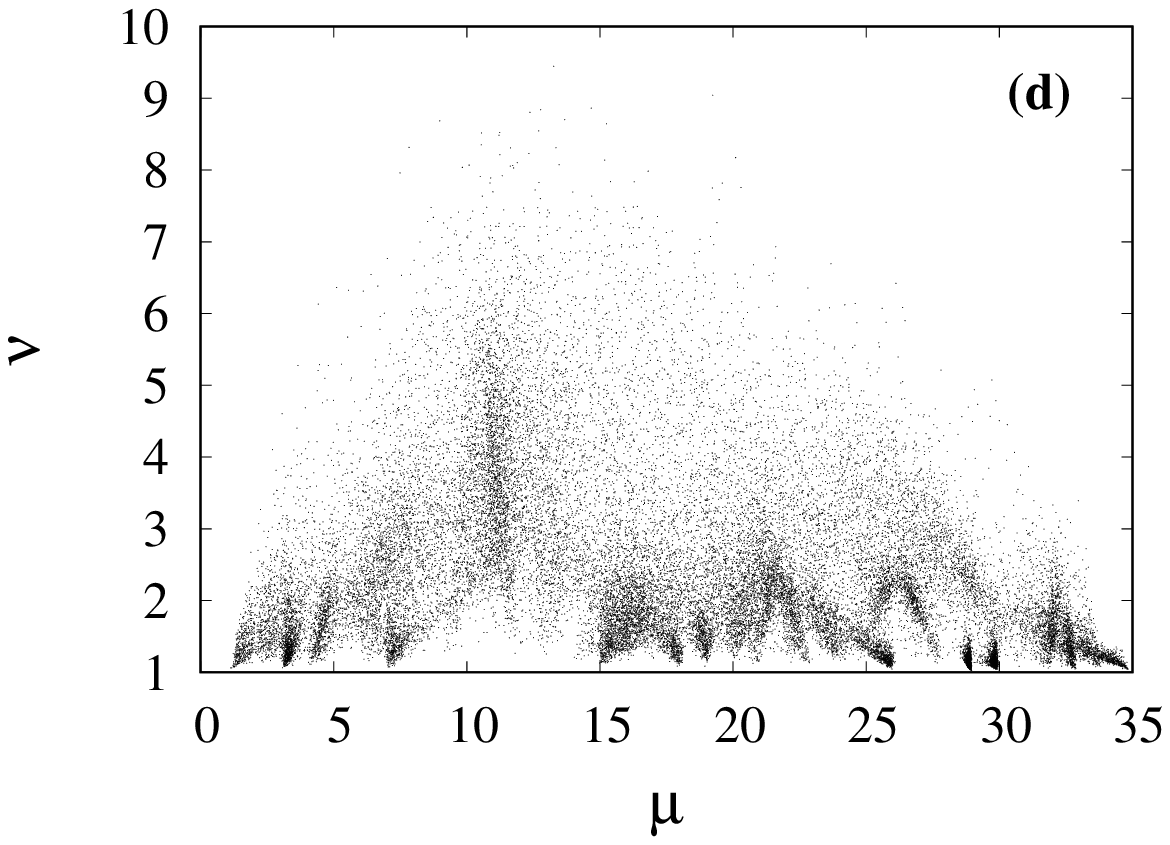}
\includegraphics[width=0.32\textwidth]{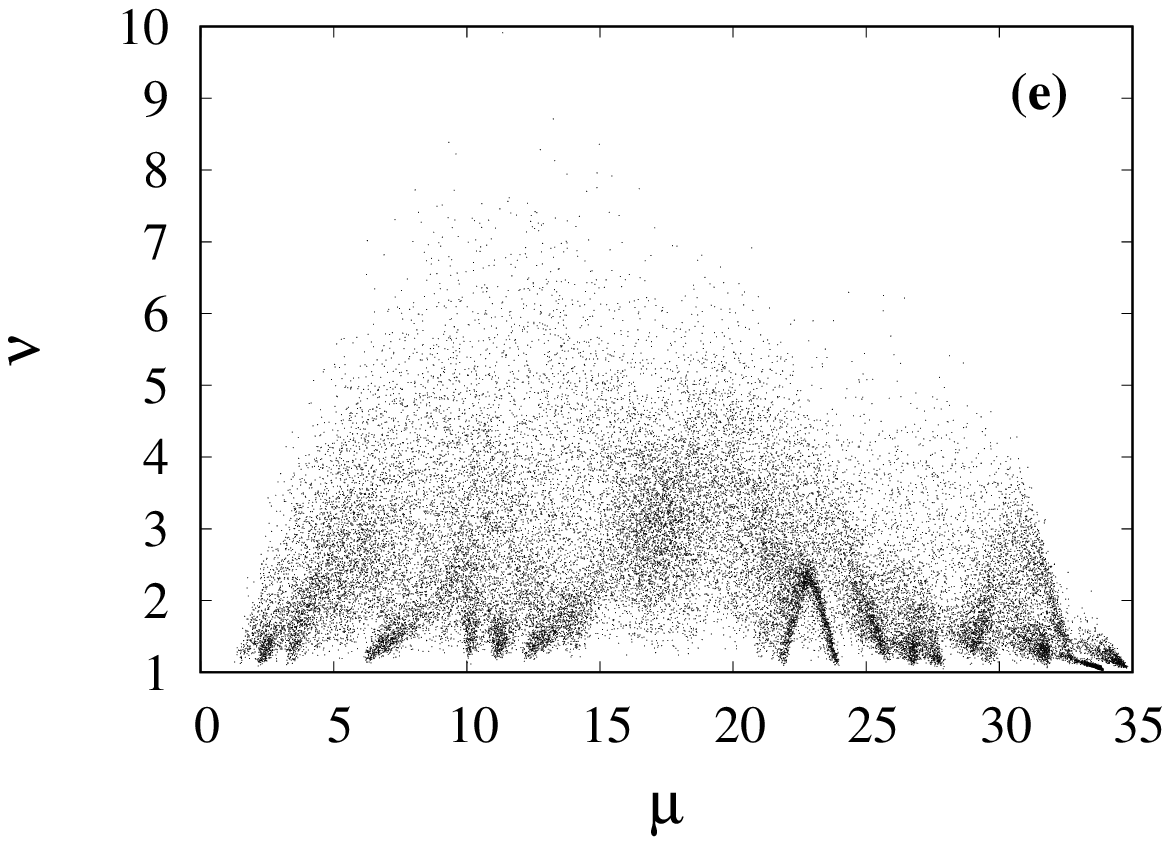}
\includegraphics[width=0.32\textwidth]{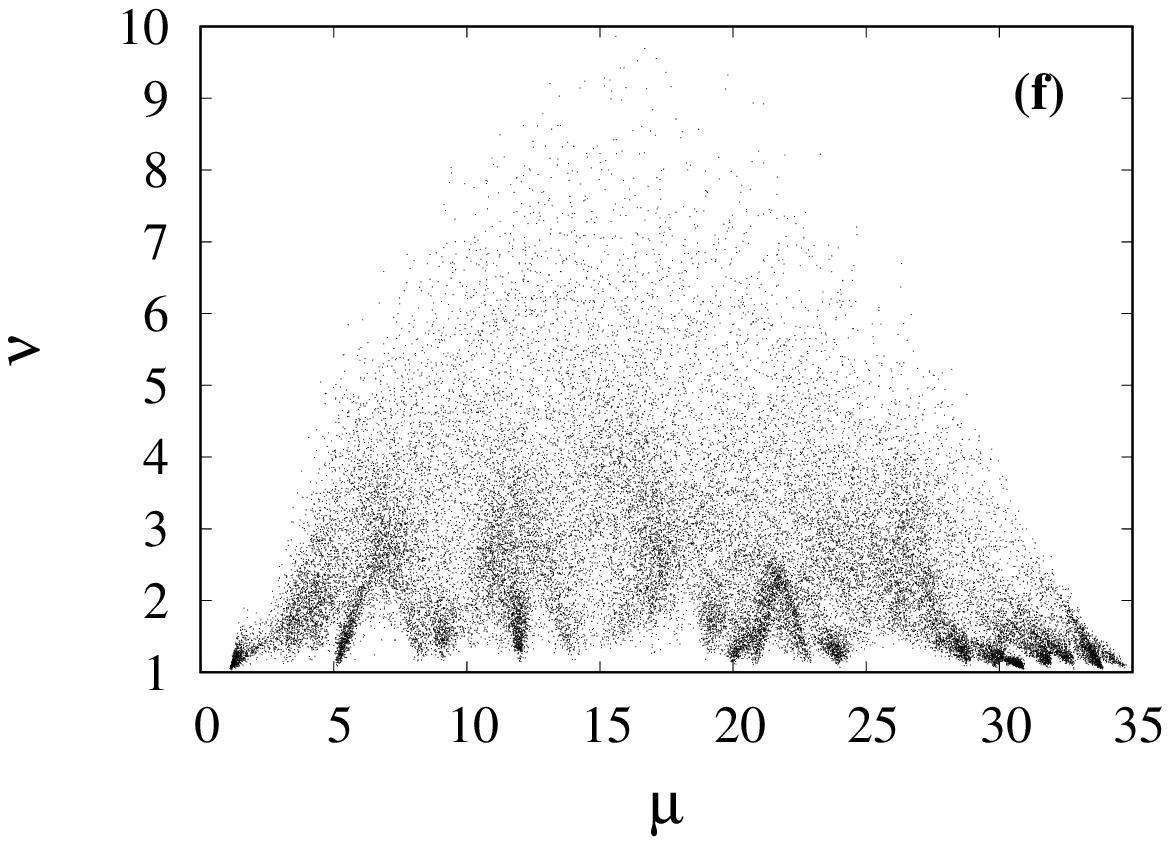}\\
\includegraphics[width=0.32\textwidth]{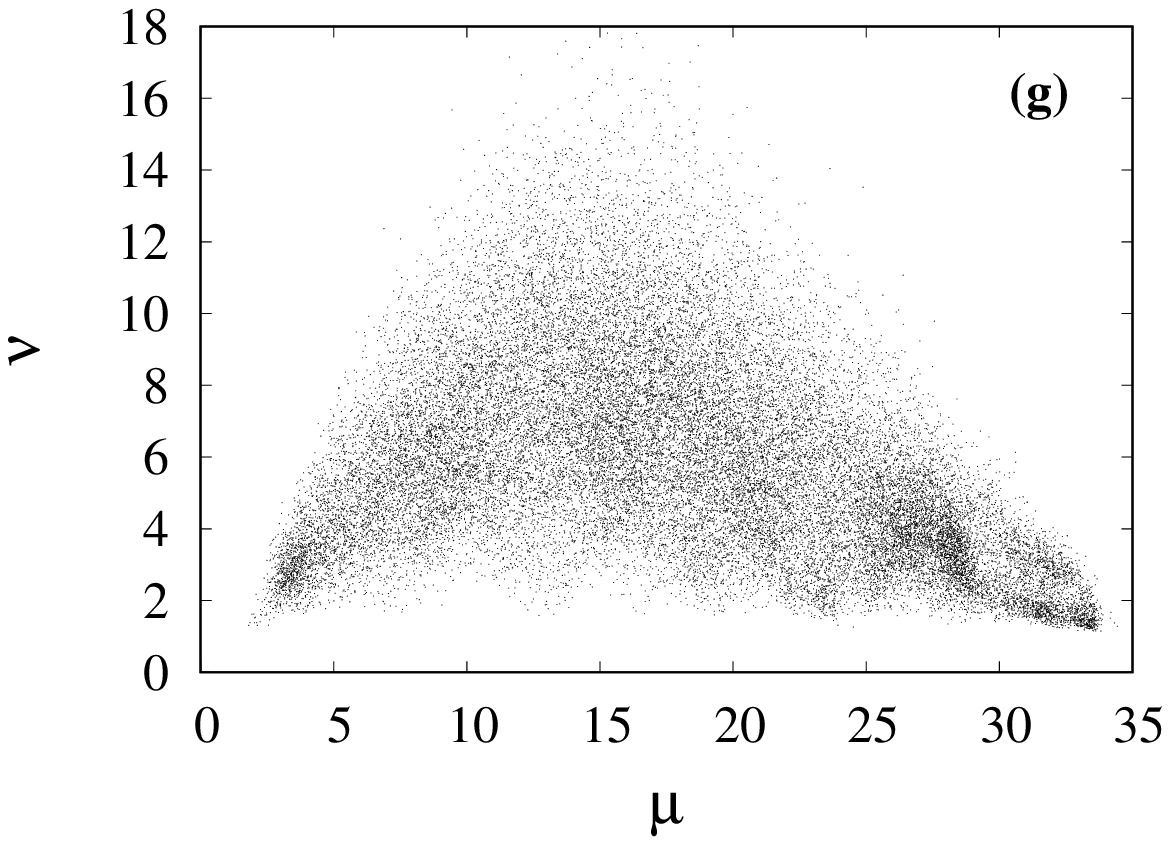}
\includegraphics[width=0.32\textwidth]{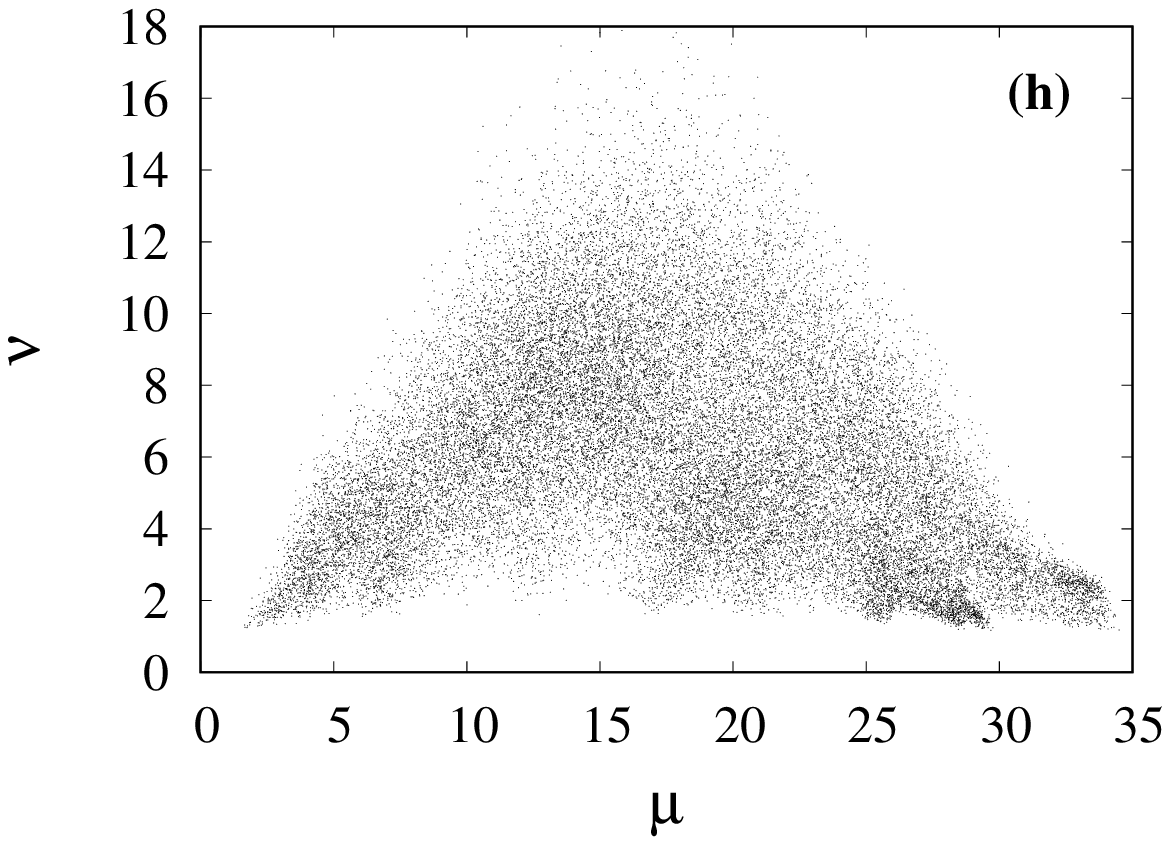}
\includegraphics[width=0.32\textwidth]{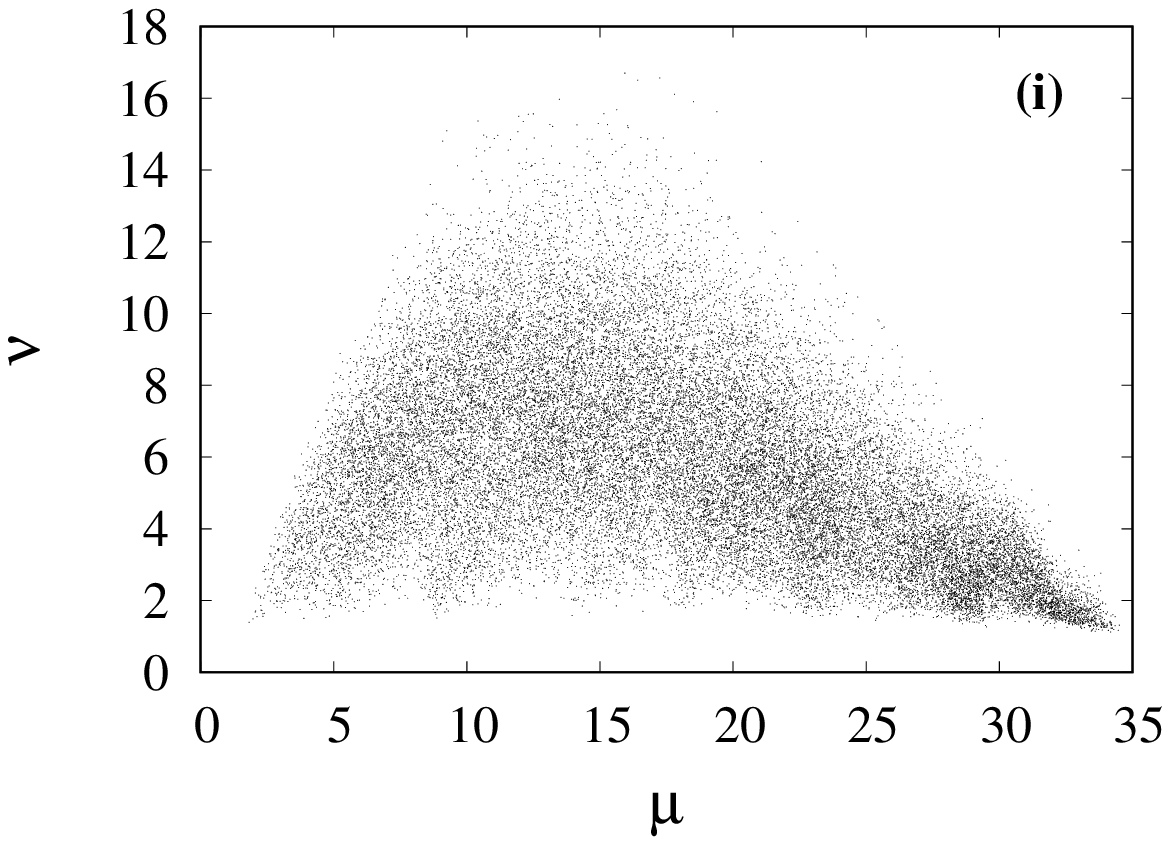}
\caption{Distribution of eigenfunctions in the $\mu$--$\nu$ plane, 
where the parameter $\mu$ is given by Eq.~(\ref{mu}), and $\nu$ is the 
participation ratio (\ref{npc}). The case of the Munk waveguide.
Distance values:
(a)-(c) $r\F=50$~km, (d)-(f) $r\F=250$~km, and (g)-(i) $r\F=1000$~km.
The sound frequency is 25~Hz. Left column corresponds to numerical solution of the parabolic equation, middle and right columns correspond
to RMT modeling with $\Delta r=10$~km and $\Delta r=50$~km, respectively.}%
\label{fig-munkfunc25}
\end{figure}

Eigenfunction distributions in the plane of parameters $\mu$ and $\nu$ allow for visualization and physical interpretation of coherent phenomena \cite{PRE87,PhysScr}.
They are depicted in Fig.~\ref{fig-bepfunc25} for the BEP model 
and Fig.~\ref{fig-munkfunc25} for the Munk model.
Each plot in Figs. \ref{fig-bepfunc25} and \ref{fig-munkfunc25} is superposition of eigenfunction distributions corresponding to 
individual realizations. 
Coherent features are seen as some structures with much higher density of plotting points as compared to fuzzy background.
For instance,
one can see that resulting distributions include vertical stripes, well-resolved or slightly smeared.
Each such stripe corresponds to a set of eigenfunctions having nearly the same value of the identifier $\mu$.
There are two kinds of the stripes, differing by interval of $\nu$ they belong to.
The stripes adjoining to the minimally accessible value $\nu=1$ are formed by eigenfunctions that almost coincide 
with normal modes of the background waveguide. They are especially pronounced in the plots corresponding to $r\F=50$~km.

There are also long vertical slightly smeared stripes that don't adjoin $\nu=1$. 
They belong the interval $2\le \nu \le 6$ for the BEC model, 
and the interval $2\le \nu \le 8$ for the Munk waveguide.
These stripes correspond
to the phenomenon that was firstly described in Ref.~\cite{Viro99} and referred to as mode-medium resonance.
Mode-medium resonance means coherent dynamics of some compact group of modes.
Onset of such group is associated with the principle of ray-mode duality,
i.~e. projection of underlying ray motion onto wave dynamics in the modal space.
Particularly, mode-medium resonance  for the propagator $\hat G(r\F)$
is a wave manifestation of the ray-medium resonance that arises in the one-step Poincar\'e map \cite{PRE87,PRE73} being the ray counterpart of the propagator $\hat G$.
Ray-medium resonance corresponds to the condition 
\begin{equation}
 l_1 D(I=I_{\text{res}}) = l_2r\F,
 \label{rescond}
\end{equation}
where $l_1$ and $l_2$ are integers, $D$ is ray cycle length (ray double loop), $I$ is ray action determined by 
\begin{equation}
 I=\frac{1}{2\pi}\oint\tan\chi(z)\,dz,
 \label{action}
\end{equation}
and $\chi(z)$ is ray grazing angle. The integration in Eq.~(\ref{action}) goes over one cycle of a ray trajectory in a background
waveguide with $\delta c=0$. 
As long as the final range $r\F$ can be chosen arbitrarily, we can call resonance (\ref{rescond}) as {\it quasi-resonance}.

In the case of the BEP model, we can write down exact expressions for $D$ and $I$:
\begin{equation}
D = 
\begin{cases}
\frac{2\pi}{a\sqrt{b^2\kappa^2-2E}},\quad E\le \frac{b^2}{2}(1-\kappa)^2,\\
\frac{\Xi+\pi}{a\sqrt{b^2\kappa^2-2E}}, \quad E> \frac{b^2}{2}(1-\kappa)^2,
\end{cases}
\label{Dbep}
\end{equation}
\begin{equation}
I=\begin{cases}
\frac{b\kappa}{a}\left(
1-\sqrt{1-\frac{2E}{b^2\kappa^2}}
\right)
\\
\frac{b}{a}
\biggl(\biggr.\frac{\kappa}{2}-
\frac{\kappa}{\pi}\arcsin{\dfrac{\kappa-1}{Y}}-
\frac{\pi+\Xi}{2\pi}
\sqrt{\kappa^2-Y^2}
\biggl.\biggr)+\frac{|p(z=0)|}{\pi a},
  \end{cases}
\label{Ibep}
\end{equation}
where
\begin{equation}
 E = \frac{\tan^2\chi}{2} + U(z)
 \label{E}
 \end{equation}
 is invariant in the range-independent waveguide,
 \begin{equation}
 Y = \frac{\sqrt{2E}}{b},\quad 
 \Xi = 2\arcsin\left(\frac{\kappa-\kappa^2+Y^2}{Y} \right).
 \label{yxi}
\end{equation}
Ray action (\ref{action})
allows one to establish ray-mode correspondence by means of the Einstein-Brillouin-Keller (EBK)
quantization rule that is
\begin{equation}
 k_0I_m = m - 1/2 
 \label{EBK1}
\end{equation}
for purely-water rays, and
\begin{equation}
 k_0I_m = m - 1/4
 \label{EBK2}
\end{equation}
for rays reflecting from the ocean surface. 
Replacing $m$ by $\mu$ in Eqs.~(\ref{EBK1}) and (\ref{EBK2}), and using
 Eqs.~(\ref{Dbep})-(\ref{yxi}), one can identify quasi-resonances responsible for vertical stripes for the BEP model.
 In the case of the Munk waveguide, dependence $D(I)$ can be found numerically \cite{Smirnov01}.
 So, one can find out that the plots corresponding to $r\F=50$~km depict mode-medium  quasi-resonance with $l_1=l_2=1$. 
Modes corresponding to this quasi-resonance obey to Eq.~(\ref{EBK1}) in the case 
of the Munk model, and to Eq.~(\ref{EBK2}) in the case of the BEP model.

Rays belonging to neighbourhood of the resonant action $I_{\text{res}}$ corresponding to some pair ($l_1$,$l_2$)
 form a cluster that maintains coherence until this quasi-resonance is well-isolated in the action space from quasi-resonances
with other values of $l_1$, $l_2$ and $I_{\text{res}}$.
Eqs.~(\ref{EBK1}) and (\ref{EBK2}) allow one to project the cluster onto normal modes.
Thus, the relating normal modes obey inequality
\begin{displaymath}
 m_{\text{res}} - \Delta m\le m \le m_{\text{res}} + \Delta m,
\end{displaymath}
where $m_{\text{res}}\simeq k_0I_{\text{res}}$ is mode number corresponding to the exact ray-medium quasi-resonance (\ref{rescond}), 
and $\Delta m$ is related to half-width of ray-medium quasi-resonance
$\Delta I$ as $\Delta m=k_0\Delta I$.
Rays compounding a resonance-induced cluster have close travel times, providing spontaneous wavefield focusing in the time domain \cite{Chaos,PRE73}.
This circumstance makes them observable in experiments.
As it was shown in Ref.~\cite{PRE73}, density of ray-medium quasi-resonances linearly increases with increasing $r\F$, therefore, all such clusters, as well as related modes,
eventually overlap and loss their coherence.

It is worthwhile to mention that spectrum of the RMT-based propagator with $\Delta r=10$~km for the BEP model
doesn't have resonance-induced vertical stripe, in contrast to the case of $\Delta r=50$~km.
In the Fig.~\ref{fig-munkfunc25}(b) corresponding to the Munk waveguide, the stripe is visible but much shorter than in Figs.~\ref{fig-munkfunc25}(a) and \ref{fig-munkfunc25}(c).
Indeed, rays corresponding to the quasi-resonance have cycle length $D$ of 50 km. It is much larger than
$\Delta r=10$~km. So, it turns out that the propagator with $\Delta r=10$~km ignores cross-mode correlations giving rise to mode-medium quasi-resonance.
It leads to the conclusion that the correct reproduction of coherent resonance-induced phenomena requires $\Delta r$ to exceed the maximal value 
of cycle length for rays propagating without reaching the bottom.
This rule, however, doesn't apply if mode-medium quasi-resonance losses its stability already for $\Delta r<D$. It may take place for relatively high frequencies,
in the presence of scattering on fine-scale inhomogeneities \cite{UFN}.

Another noticeable feature of eigenfunction distributions shown in Figs.~\ref{fig-bepfunc25} and \ref{fig-munkfunc25} is the presence of well-resolved curved lines
having form of bridges connecting points $(\mu=\mu',\nu=1)$ and $(\mu=\mu'',\nu=1)$. As it was shown in Ref.~\cite{PRE87}, these ``bridges''
correspond to resonant inter-mode transitions obeying
\begin{equation}
 k_0(E_m-E_n) = \frac{2\pi l}{r\F},\quad
 m>n,\quad
 l=1,2,3,\text{...}
 \label{rescond2}
\end{equation}
Complete well-drawn ``bridges'' appear if resonance (\ref{rescond2}) is well-isolated, and the corresponding eigenfunction of the propagator
can be fairly represented as sum
\begin{equation}
 \Phi_j \simeq  C_{mj}\psi_m + C_{nj}\psi_n.
 \end{equation}
Then normalization condition for eigenfunctions 
\begin{displaymath}
\int |\Phi_j(z)|^2\,dz=1
\end{displaymath}
yields
\begin{equation}
 |C_{mj}|^2 + |C_{nj}|^2 = 1.
\end{equation}
Rewriting $C_{mj}$ and $C_{nj}$ as
\begin{displaymath}
 C_{mj} = \cos\vartheta,\quad 
 C_{nj} = \sin\vartheta,
\end{displaymath}
we obtain
\begin{equation}
 \mu = m\cos^2\vartheta + n\sin^2\vartheta,\quad
 \nu = \frac{1}{\cos^4{\vartheta} + \sin^4{\vartheta}}.
 \end{equation}
Basically $\vartheta$ is a random quantity and depends on realization of the sound-speed inhomogeneity.
Superposition of large number of propagator realizations uniformly fills full range of $\vartheta$ values, from $-\pi$ to $\pi$. 
It results in drawing the parametric curve $\mu(\vartheta), \nu(\vartheta)$
as a bridge-like pattern. If resonance (\ref{rescond2}) is poorly isolated, the bridge-like pattern becomes distorted or even completely dissolved. 
The latter eventually happens with increasing of $r\F$ due to multiplication of triplets $(l,m,n)$ satisfying Eq.~(\ref{rescond2}).

In general, appearance of resonance-assisted features is linked to the aforementioned spectral equivalence of the propagator $\hat G$ and 
the  Floquet operator for the fictitious waveguide with sound speed inhomogeneity given by Eq.~(\ref{fiction}).
As long as positions of quasi-resonances in the modal space depend on $r\F$, we conclude that the resonance-induced structures 
don't mean enhanced transfer between certain modes.
Their actual meaning is indication of ability of related modal groups to maintain coherence for $r\le r\F$.
Presence of such structures is an unambiguous signature of long-living cross-mode correlations.
It is worth emphasizing that such structures are also observed in the plots obtained using the RMT approach.
So, it turns out RMT is able to reproduce quite subtle coherent effects.
This allows one to classify the RMT approach as a really powerful method of modeling.

As $r\F$ increases number of quasi-resonances (\ref{rescond}) and (\ref{rescond2}) grows.
Since that isolation of individual quasi-resonances is violated. 
Overlapping of quasi-resonances results in their destruction \cite{PRE73}.
So, the quasi-resonances are no more able to produce coherent structures in the $\mu$-$\nu$ plane. 
The resulting pattern acquires the form of a fuzzy spread boomerang, like in Figs.~\ref{fig-bepfunc25}(g-i) and \ref{fig-munkfunc25}(g-i).

\subsection{Signal frequency 75 Hz}

\begin{figure}[!htb]
\begin{center}
\includegraphics[width=0.47\textwidth]{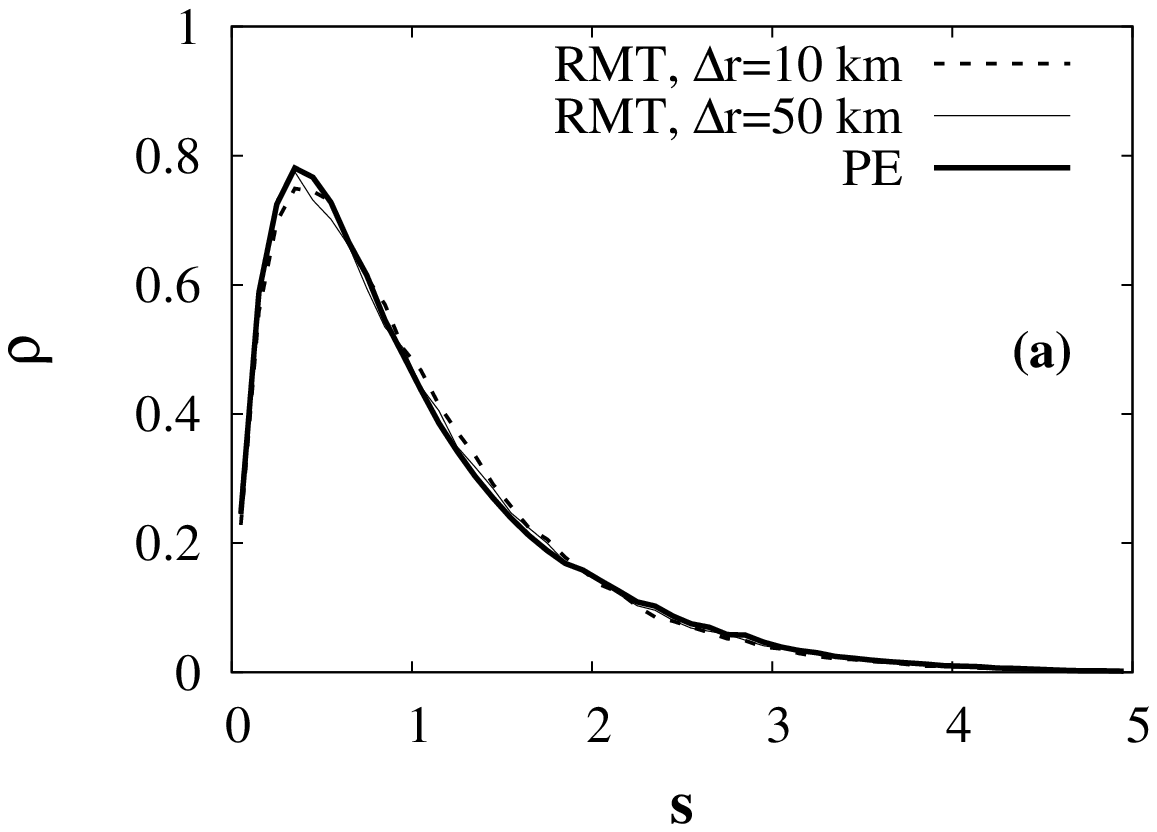}
\includegraphics[width=0.47\textwidth]{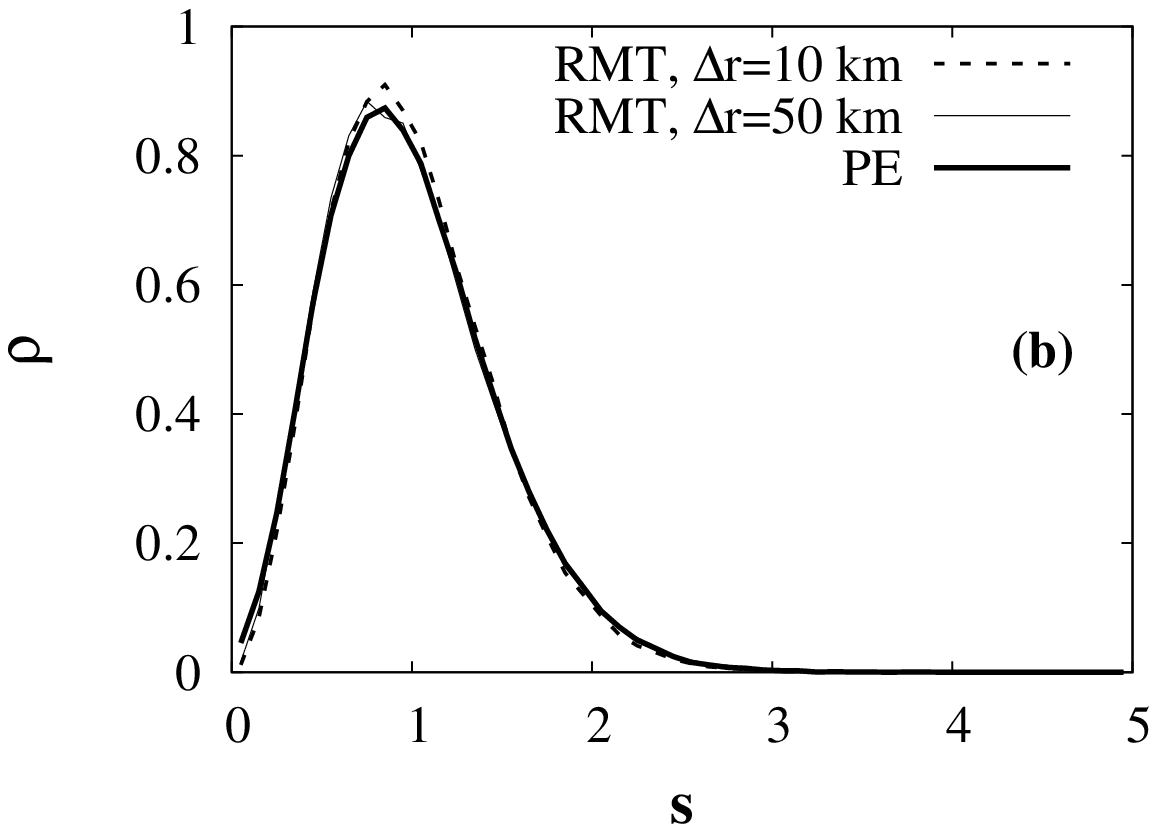}
\end{center}
\caption{Level spacing distribution for the propagator $\hat G(r\F)$ corresponding to the Munk model and
signal frequency is of 75 Hz. 
Figures (a) and (b) correspond to the $r\F=50$ and $r\F=1000$ km, respectively. }%
\label{fig-distr75}
\end{figure}
\begin{figure}[!htb]
\includegraphics[width=0.47\textwidth]{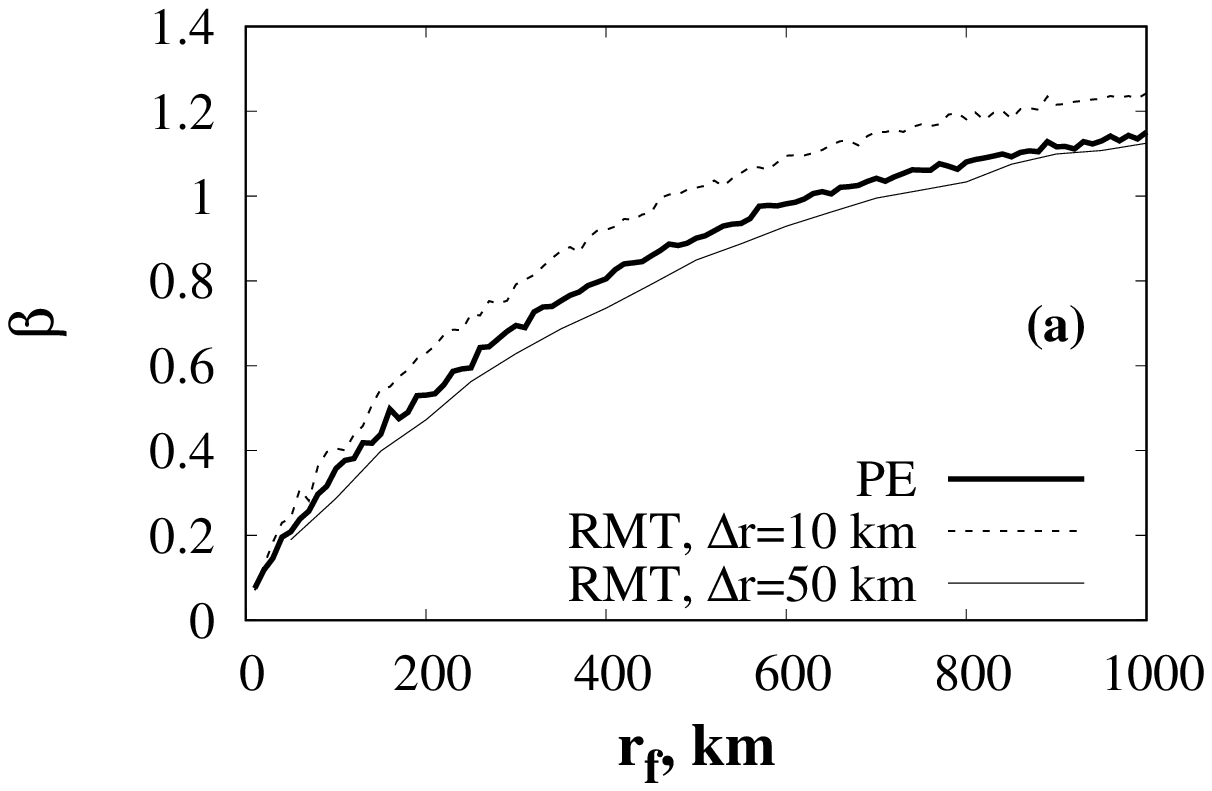}
\includegraphics[width=0.47\textwidth]{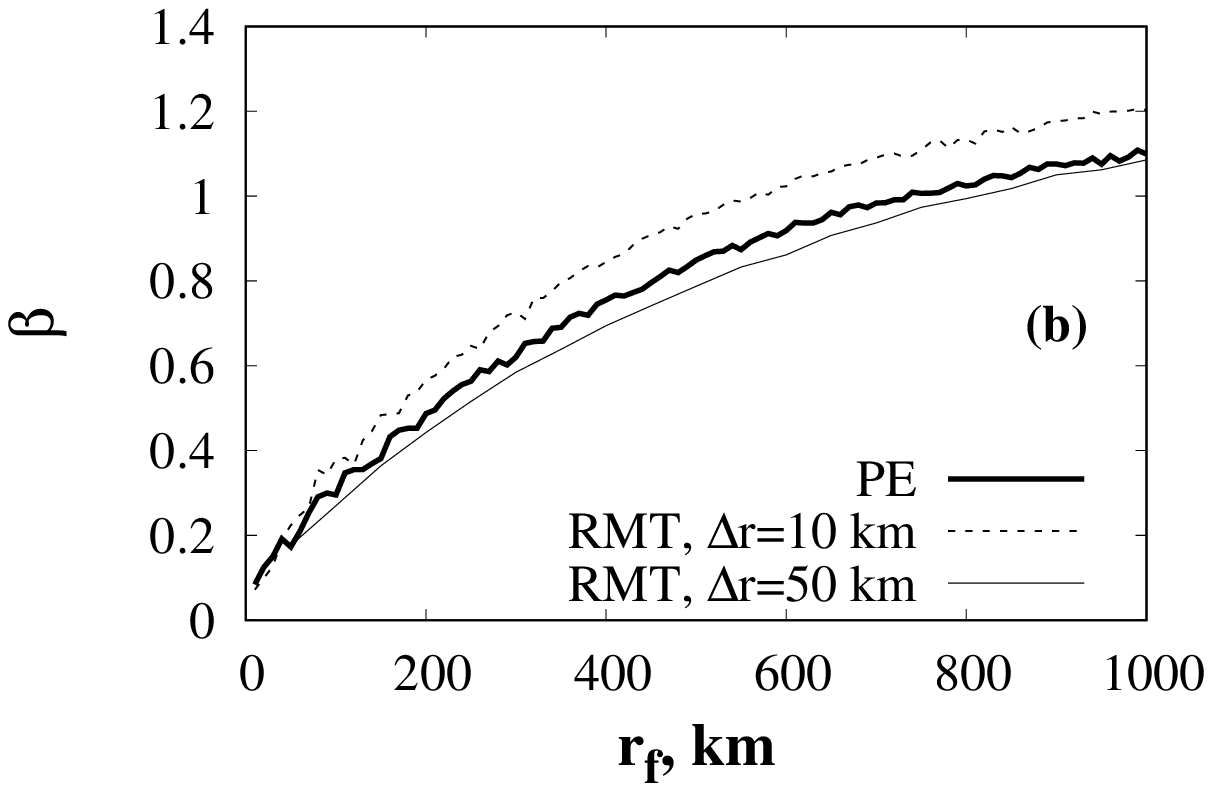}
\caption{Range dependence of the Brody parameter $\beta$ corresponding to the best fit of level spacing distribution.
Panel (a) corresponds to the BEP model, panel (b) depicts results for the canonical Munk waveguide. Signal frequency is of 75 Hz.}%
\label{fig-brody75}
\end{figure}
\begin{figure}[!htb]
\includegraphics[width=0.43\textwidth]{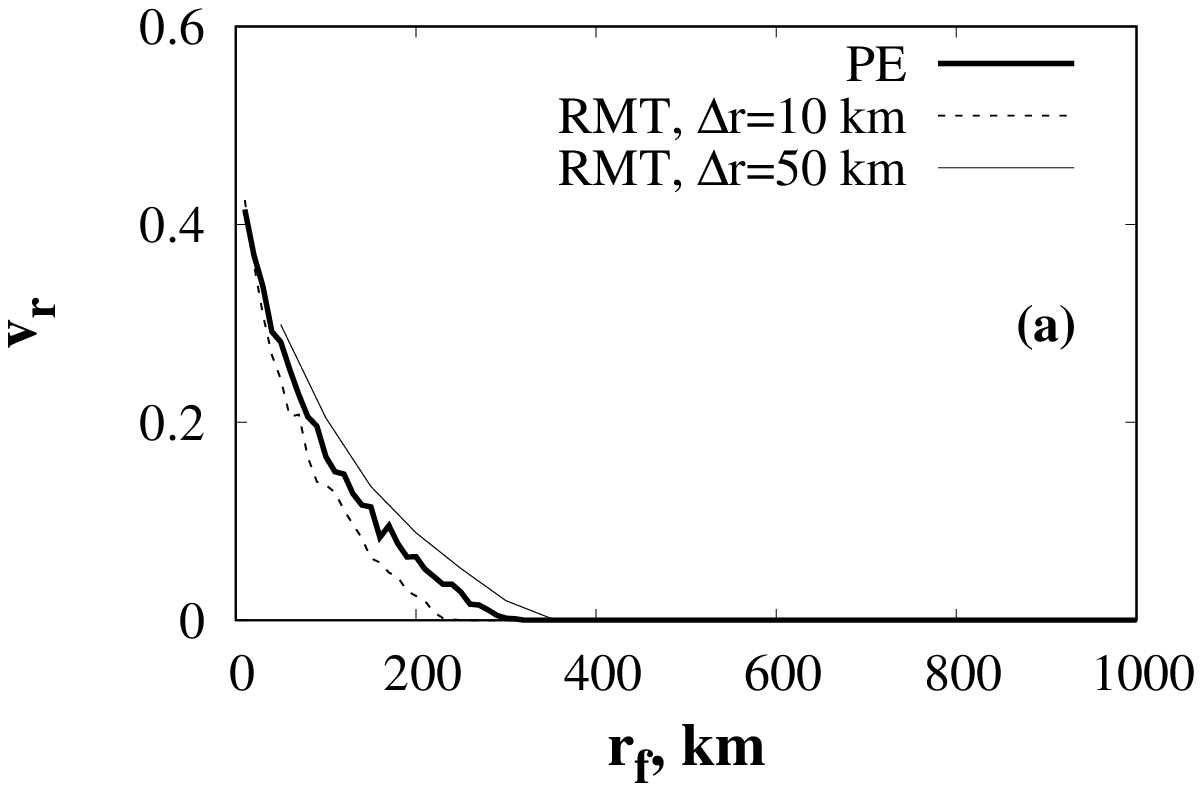}
\includegraphics[width=0.43\textwidth]{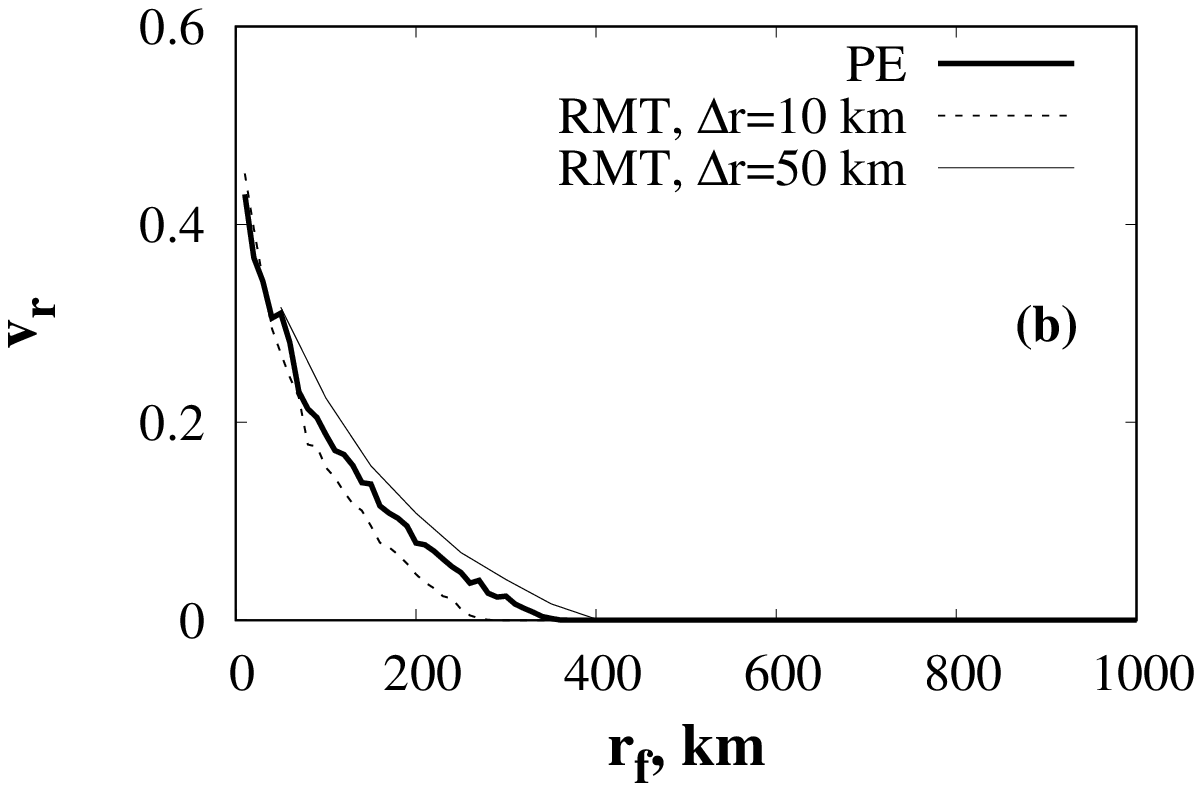}
\caption{Fraction of coherently propagating modes $v_{\text{r}}$ corresponding to the best fit of level spacing distribution by the Berry-Robnik formula (\ref{berrob}).
Panel (a) corresponds to the BEP model, panel (b) depicts results for the canonical Munk waveguide. Signal frequency is of 75 Hz.}%
\label{fig-berry75}
\end{figure}

Increasing of signal frequency substantially imrpoves correspondence between the RMT-based propagator and its directly-calculated counterpart. 
In the case of $f=75$~Hz the corresponding level spacing distributions almost fully coincide, as it is shown
in Fig.~\ref{fig-distr75}.
They all expose transition to Wigner statistics as $r\F$ grows.
It is well demonstrated by range dependence of the Brody parameter corresponding to the best fit (see Fig.~\ref{fig-brody75}).
It also indicates good agreement between
the RMT and Crank-Nicolson data. Notably, the curve corresponding to RMT data with $\Delta r=50$ km is closer to the results of direct solving that
one corresponding to $\Delta r=10$~km.
Fitting level spacing distribution by means of the Berry-Robnik formula (\ref{berrob}), we can estimate fraction of modes which don't experience strong
scattering and propagate coherently. Figure \ref{fig-berry75} shows that this fraction rapidly decreases with range and completely vanishes for $r\F$ of nearly 
400 km. 

Eigenfunction distributions in the $\mu$-$\nu$ plane exhibit signatures of mode-medium quasi-resonance  only for relatively short ranges.
RMT modeling is able to reproduce it, but with much less resolution than in the case of the direct solution.
For longer ranges, the distributions rapidly achieve the boomerang-like pattern representing a kind of statistical equillibrium.

So, we can see that coherent structures are remarkably suppressed for $f=75$~Hz, as compared to the case of $f=25$~Hz.
These drastic differences in spectral statistics indicate qualitatively different mechanismes of scattering, namely transition to ray-like
scattering with impact of ray chaos.

\begin{figure}[!htb]
\includegraphics[width=0.48\textwidth]{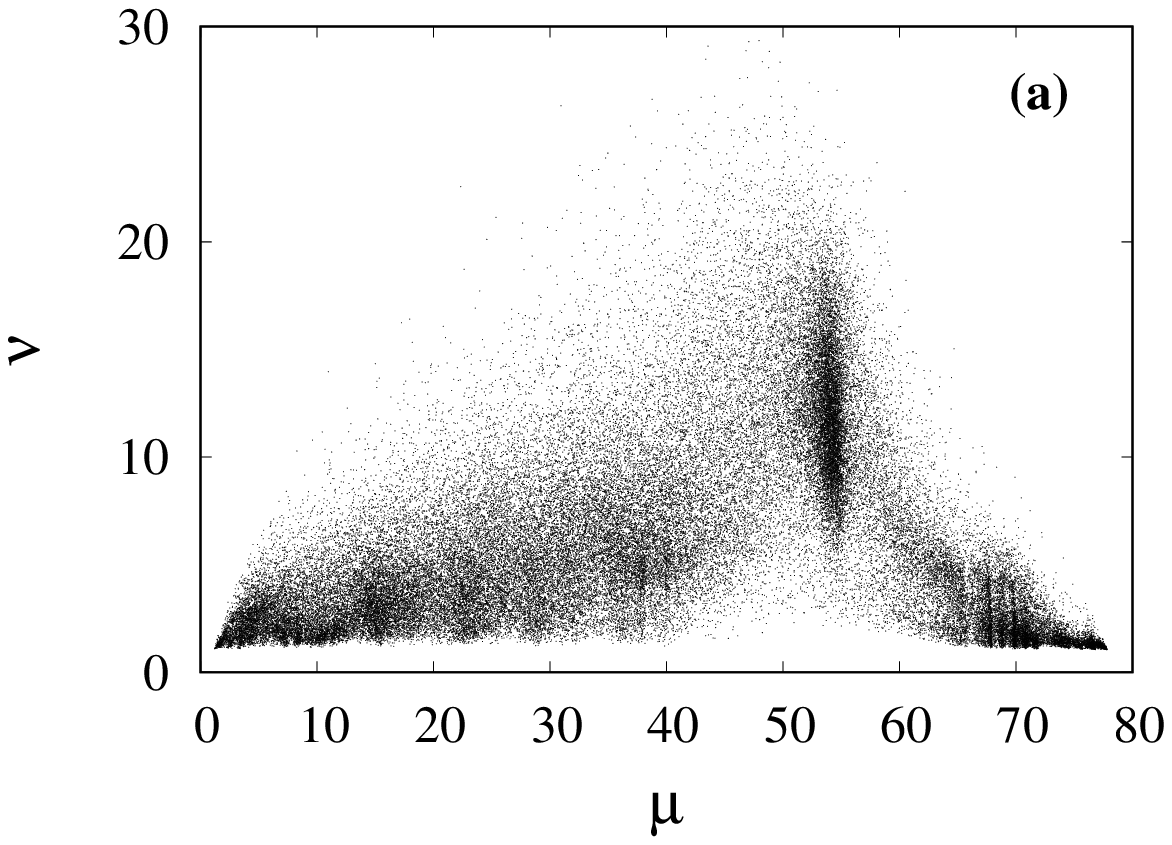}
\includegraphics[width=0.48\textwidth]{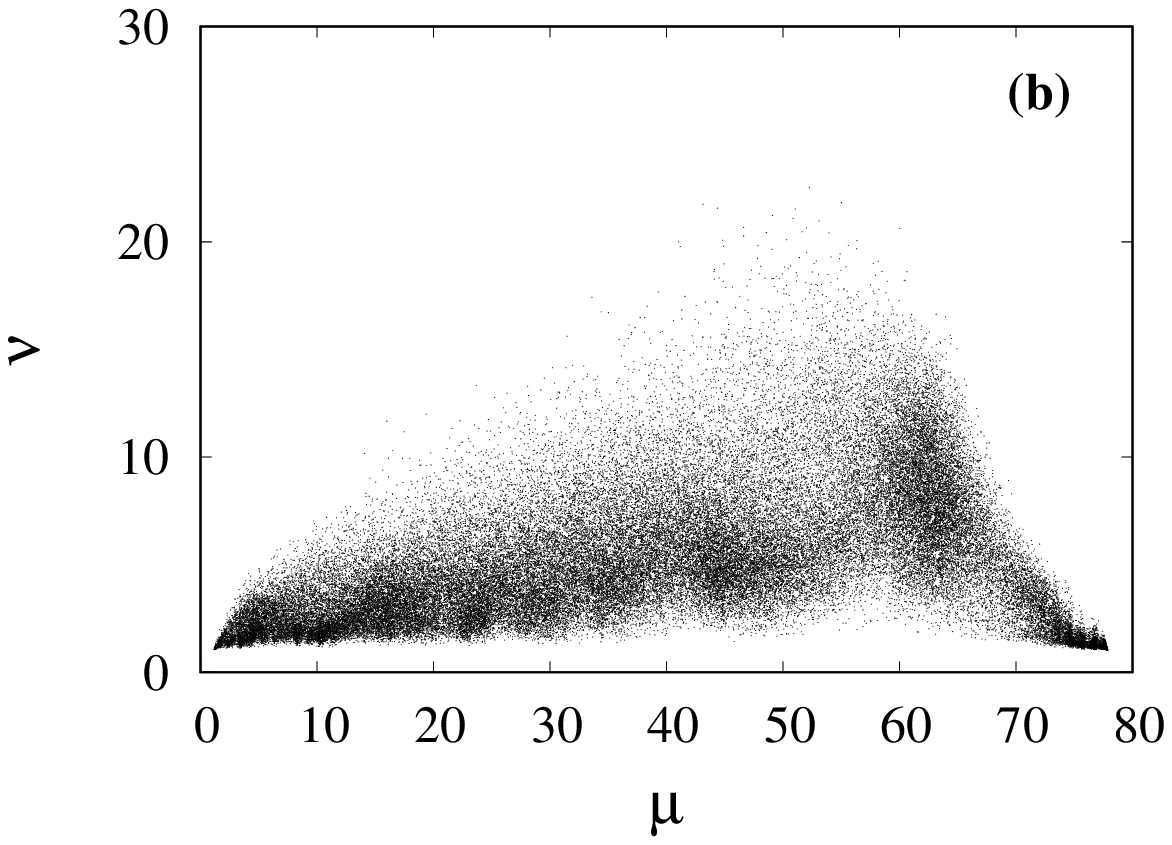}\\
\includegraphics[width=0.48\textwidth]{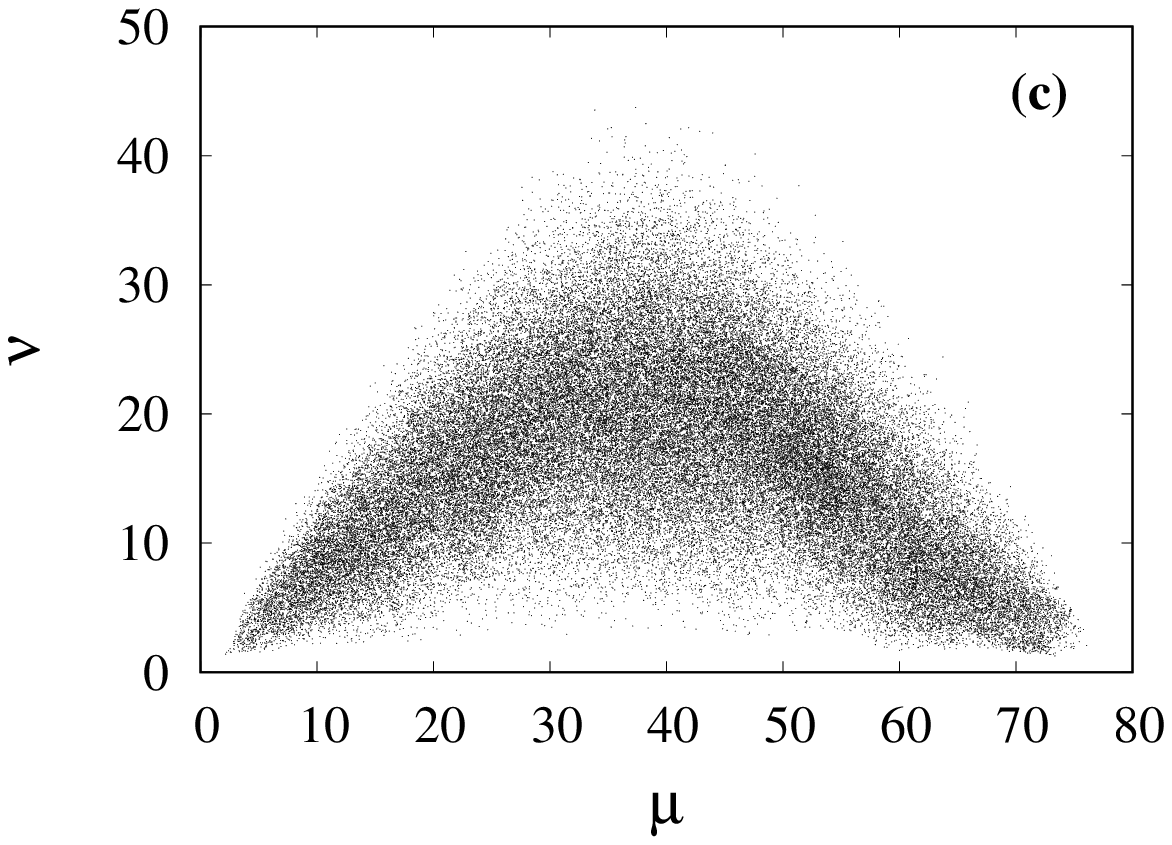}
\includegraphics[width=0.48\textwidth]{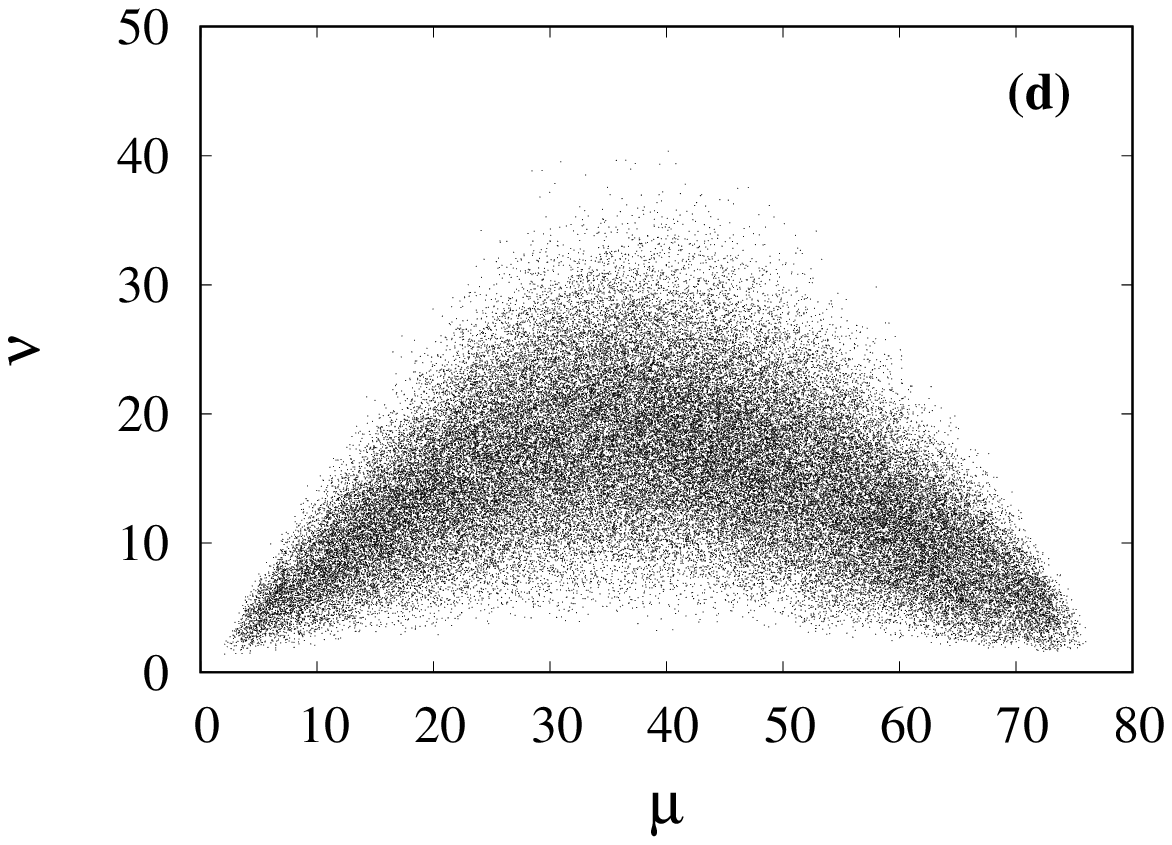}
\caption{Distribution of eigenfunctions in the $\mu$--$\nu$ plane, 
where the parameter $\mu$ is given by Eq.~(\ref{mu}), and $\nu$ is the 
participation ratio (\ref{npc}). The case of the BEP model of a waveguide.
Sound frequency is 25~Hz. 
Upper and lower raws correspond to $r\F=50$~km and $r\F=250$~km, respectively.
The left column corresponds to numerical solution of the parabolic equation, the right column depicts data
obtained via the RMT modeling with $\Delta r=50$~km.}%
\label{fig-bepfunc75}
\end{figure}
\begin{figure}[!htb]
\includegraphics[width=0.48\textwidth]{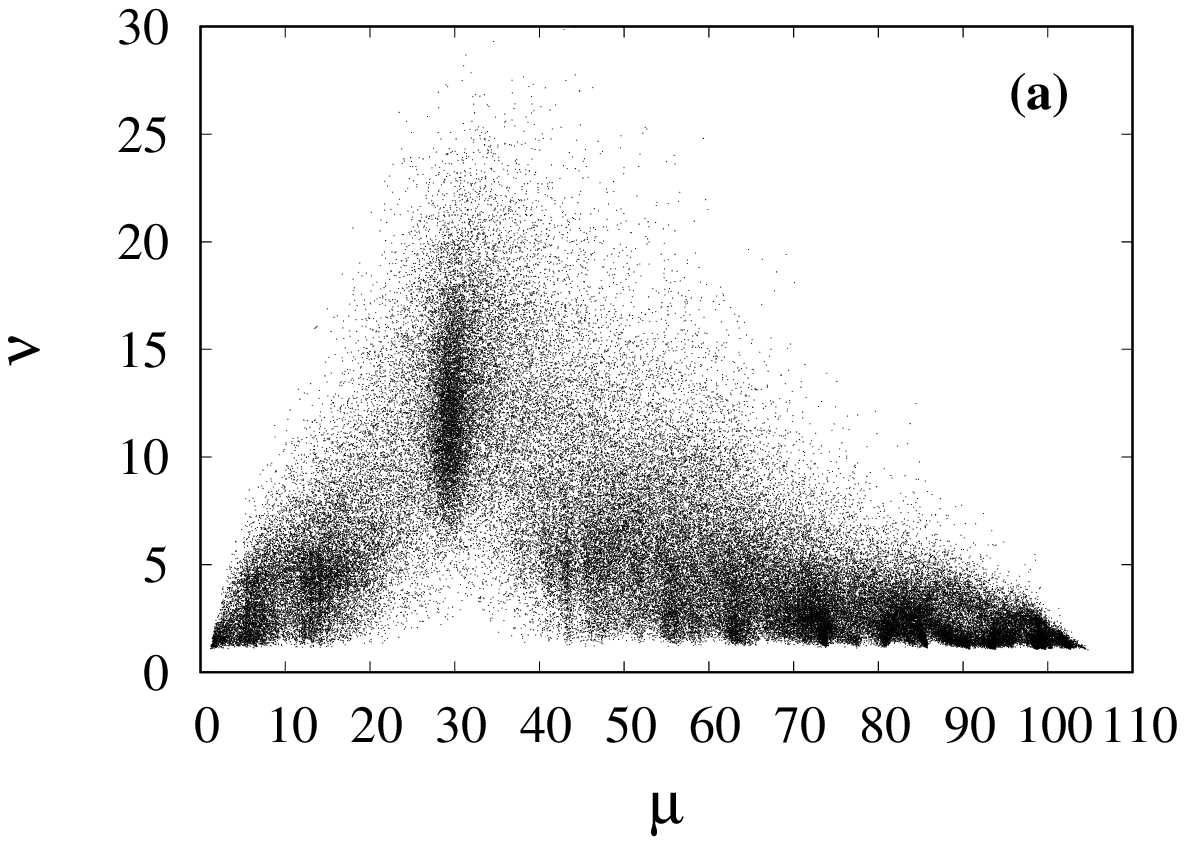}
\includegraphics[width=0.48\textwidth]{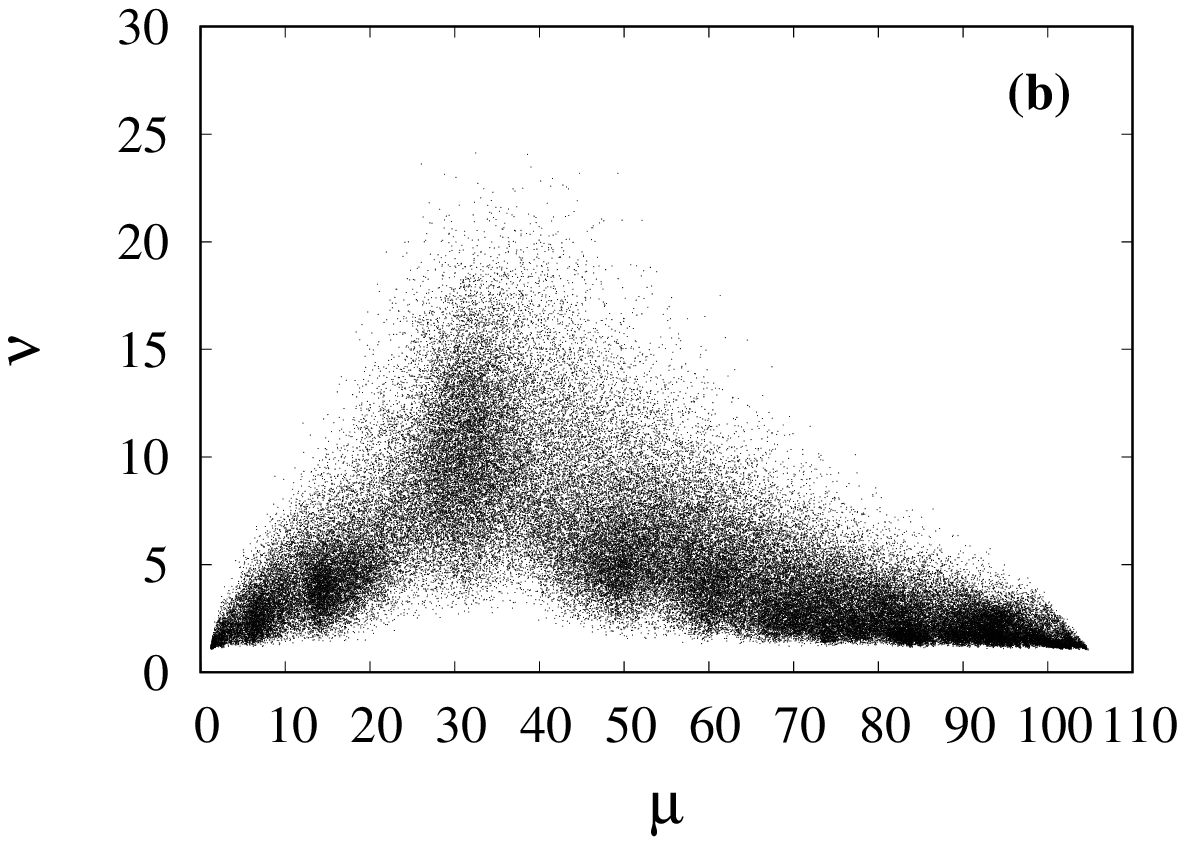}\\
\includegraphics[width=0.48\textwidth]{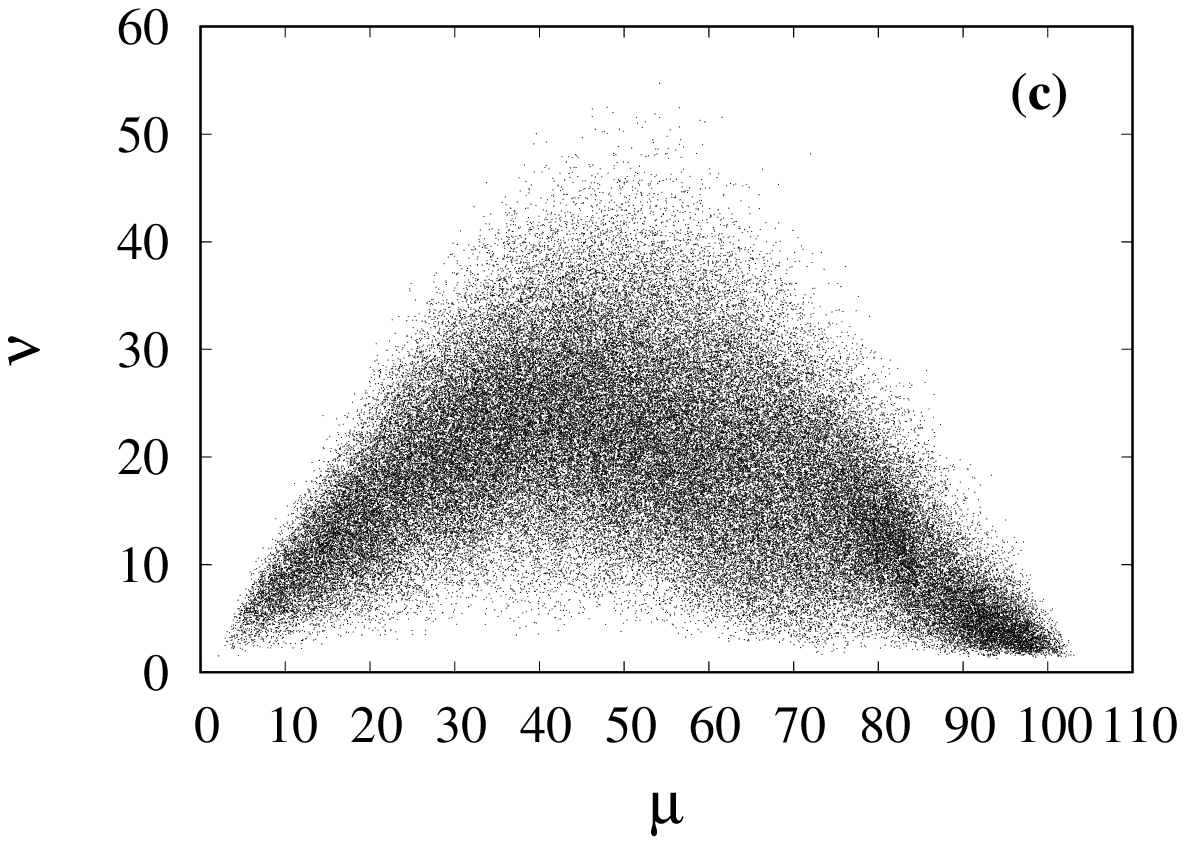}
\includegraphics[width=0.48\textwidth]{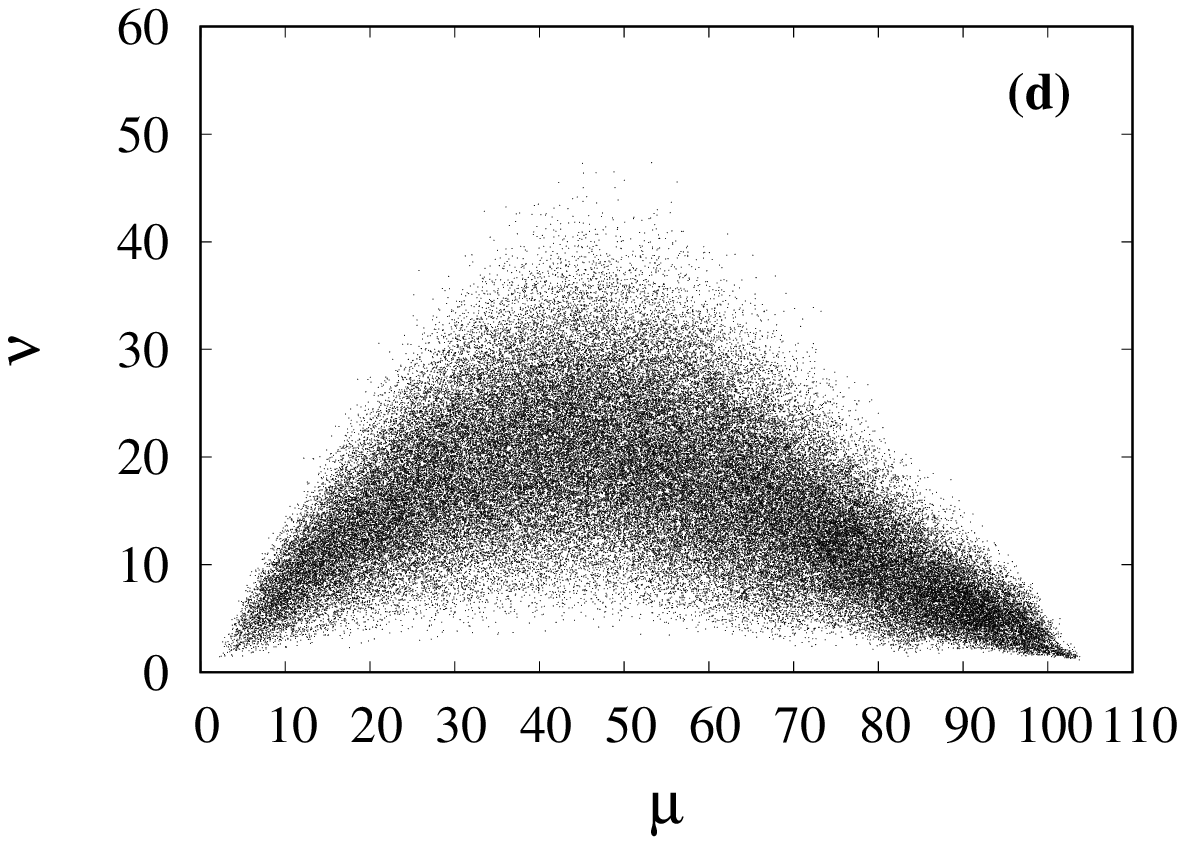}
\caption{Distribution of eigenfunctions in the $\mu$--$\nu$ plane, 
where the parameter $\mu$ is given by (\ref{mu}), and $\nu$ is the 
participation ratio (\ref{npc}). The case of the Munk waveguide.
Sound frequency is 25~Hz. 
Upper and lower raws correspond to $r\F=50$~km and $r\F=250$~km, respectively.
The left column corresponds to numerical solution of the parabolic equation, the right column depicts data
obtained via the RMT modeling with $\Delta r=50$~km.}%
\label{fig-munkfunc75}
\end{figure}

\section{Summary}

In the present paper we examine the approach based on construction of a wavefield propagator using the random matrix theory (RMT).
In particular, we study its ability to reproduce properties of the propagator constructed by means 
of direct numerical solution of the parabolic equation.
It is found out that noticeable differences between the RMT-based propagator occur only for relatively short ranges, below 100 km, where cross-mode
correlations play an essential role and give various coherent impacts to spectral statistics.
The coherent impacts are caused by quasi-resonances having a deterministic origin and
 especially pronounced for frequencies of few tens Hz.
To suppress influence of cross-mode correlations,
the range step $\Delta r$ of a RMT-based propagator should exceed the maximal cycle length of rays propagating without reflections off the bottom.
For longer ranges, the random matrix theory provides excellent agreement with direct solutions of the parabolic equation.
Calculations of mode energies show that the RMT-based method slightly underestimates mode coupling, but the differences with the direct solutions
are observed only for the lowest modes.

The main physical result of the present work is
the evident influence of ray-based features
onto propagator spectra for low signal frequencies.
Somewhat unexpectedly, they are much more revealed in the propagator spectrum at the frequency 
of 25 Hz than at the frequency of 75 Hz.
Mode-medium quasi-resonances, being modal counterparts of ray-medium quasi-resonances, 
are responsible for specific vertical stripes in distributions of propagator eigenfucntions
in the plane with parameters $\mu$ and $\nu$ as coordinates.
In the ray picture, the corresponding ray-medium quasi-resonances are mainly destroyed due to ray chaos. 
Appearance of resonance-assisted vertical stripes in the eigenfunction statistics for low frequencies indicates
that stability of the corresponding mode-medium quasi-resonances is restored. 
It is closely related to the recovery of ray stability reported in Ref.~\cite{PRE76,Hege}.
One of such mechanismes is ray scattering on fine-scale distortions of a sound-speed profile \cite{Akust07},
which are also present in the models considered in the present paper.
For low frequencies such distortions become irrelevant for wave dynamics.
Therefore, refraction of a wavefield becomes meaningfully stabilized, as compared to the high-frequency regime.
Stabilization of refraction is reflected in dynamics of modal amplitudes \cite{Akust08} and results in recovery of mode-medium resonance.

In the present paper we consider only relatively low frequencies. In the case of higher frequencies we can expect qualitatively different properties 
of sound scattering, for example, more significant influence of fine-scale structures in the sound-speed profile. It can be reflected in
spectral statistics of a wavefield propagator.
  Another important issue is applicability of the RMT-based approach in the case of waveguides with a slowly-varying background sound-speed profile.
Also, it is reasonable to extend applicability of the RMT approach onto shallow-sea propagation, when the unitarity of the propagator doesn't hold.
All these issues will be addressed in forthcoming works.

\section*{Acknowledgments}

This work was supported by the Russian Foundation of Basic Research within the projects 16-35-60040 and 16-05-01074, and  by 
the POI FEBRAS Program
'Mathematical simulation and analysis of dynamical processes in the ocean'
(№117030110034-7)
Author is grateful to Steven Tomsovic for stimulating and fruitful discussions, and to Leonid Kon'kov for the assistance in preparation of figures.


\begin{thebibliography}{50}
%
\bibitem{AET} F. J. Beron-Vera, M. G. Brown, J. A. Colosi, S. Tomsovic, A. L. Virovlyansky,
M. A. Wolfson and G. M. Zaslavsky, 
{\it J. Acoust. Soc. Amer.} {\bf 114} (2003) 1226.
%
\bibitem{RayWave} D. Makarov, S. Prants, A. Virovlyansky and G. Zaslavsky, 
{\it Ray and Wave Chaos in Ocean Acoustics: Chaos in Waveguides} 
(World Scientific, Singapore, 2010).
%
\bibitem{UFN} A. L. Virovlyansky, D. V. Makarov and S. V. Prants, 
{\it Phys. Usp.} {\bf 55} (2012) 18.
%
\bibitem{ColosiBook} J. A. Colosi, 
{\it Sound Propagation through the Stochastic Ocean} 
(Cambridge University Press, Cambridge, 2016).
%
\bibitem{Dozier_Tappert} L. B. Dozier and F. D. Tappert, 
{\it J. Acoust. Soc. Amer.} {\bf 63} (1978) 353.
%
\bibitem{ColosiMorozov} J. A. Colosi and A.K.Morozov, 
{\it J. Acoust. Soc. Amer.} {\bf 126} (2009) 1026.
%
\bibitem{CCVO} J. A. Colosi, T. K. Chandrayadula, A. G. Voronovich and V. E. Ostashev, 
{\it J. Acoust. Soc. Amer.} {\bf 134} (2013) 3119.
%
\bibitem{Viro-WRCM16} A. L. Virovlyansky and A. Yu. Kazarova, 
{\it Waves Rand. Complex Media} {\bf 26} (2016) 564.
%
\bibitem{Hege-EPL}
K. C. Hegewisch and S. Tomsovic, 
{\it Europhys. Lett.} {\bf 97} (2012) 34002.
%
\bibitem{Hege-JASA13} K. C. Hegewisch and S. Tomsovic, 
{\it J. Acoust. Soc. Amer.} {\bf 134} Pt. 2 (2013) 3174.
%
\bibitem{Chaos} D. V. Makarov, M. Yu. Uleysky and S. V. Prants,
{\it Chaos} {\bf 14} (2004) 79.
%
\bibitem{PRE87} D. V. Makarov, L. E. Kon'kov, M. Yu. Uleysky and P. S. Petrov,
{\it Phys. Rev. E} {\bf 87} (2013) 012911.
%
\bibitem{Froufe-Perez} L. S. Froufe-Perez, M. Yepez, P. A. Mello and J. J. Saenz,
{\it Phys. Rev. E} {\bf 75} (2007) 031113.
%
\bibitem{Stockman} H.-J. St\"ockmann, 
{\it Quantum Chaos: an Introduction} 
(Cambridge University Press, Cambridge, 2007).
%
\bibitem{Viro05}
I. P. Smirnov, A. L. Virovlyansky, M. Edelman and G. M. Zaslavsky,
{\it Phys. Rev. E} {\bf 72} (2005) 026206.
%
\bibitem{PRE76} L. E. Kon'kov, D. V. Makarov, E. V. Sosedko and M. Yu. Uleysky,
{\it Phys. Rev. E} {\bf 76} (2007) 056212.
%
\bibitem{SibFU} D. V. Makarov, L. E. Kon'kov and M. Yu. Uleysky,
{\it Journ. Siber. Fed. Univ. Math. Phys.} {\bf 3} (2010) 336.
%
\bibitem{Kol97} A. R. Kolovsky,
{\it Phys. Rev. E} {\bf 56} (1997) 2261.
%
\bibitem{BR} M. V. Berry and M. Robnik,
{\it J. Phys. A: Math. Gen.} {\bf 17} (1984) 2413.
%
\bibitem{Prosen} T. Prosen,
{\it J. Phys. A: Math. Gen.} {\bf 31} (1998) 7023.
%
\bibitem{DAN} D. V. Makarov, S. V. Prants and M. Yu. Uleysky,
{\it Doklady Earth Sci.} {\bf 382} (2002) 106.
%
\bibitem{ColBr} J. A. Colosi and M. G. Brown, 
{\it J. Acoust. Soc. Amer.} {\bf 103} (1998) 2232.
%
\bibitem{Radiophys}
D. V. Makarov, L. E. Kon'kov and P. S. Petrov,
{\it Radiophys. Quant. Electron.} {\bf 59} (2016) 576.
%
\bibitem{LeBlanc} L. R. LeBlanc and F. H.  Middleton,
{\it J. Acoust. Soc. Amer.} {\bf 67} (1980) 2055.
%
\bibitem{PhysScr} D. V. Makarov and L. E. Kon'kov,
{\it Phys. Scr.} {\bf 90} (2015) 035204.
%
\bibitem{Viro99}
A. L. Virovlyansky and G. M. Zaslavsky,
{\it Phys. Rev. E} {\bf 59} (1999) 1656.
%
\bibitem{PRE73}
D. V. Makarov, M. Yu. Uleysky, M. V. Budyansky and S. V. Prants,
{\it Phys. Rev. E} {\bf 73} (2006) 066210.
%
\bibitem{Smirnov01}
I. P. Smirnov, A. L. Virovlyansky and G. M. Zaslavsky,
{\it Phys. Rev. E} {\bf 64} (2001) 036221.
%
\bibitem{Hege}
K. C. Hegewisch, N. R. Cerruti and S. Tomsovic, 
{\it J. Acoust. Soc. Amer.} {\bf 117} (2005) 1582.
%
\bibitem{Akust07} D. V. Makarov and M. Yu. Uleysky,
{\it Acoust. Phys.} {\bf 53} (2007) 495.
%
\bibitem{Akust08} D. V. Makarov, L. E. Kon'kov and M. Yu. Uleysky,
{\it Acoust. Phys.} {\bf 54} (2008) 382.







\end{thebibliography}
\end{document}